\documentclass[aps,prd,twocolumn,amsmath,showpacs,superscriptaddress,nofootinbib,nopreprintnumbers]{revtex4-1}

\usepackage{verbatim}
\usepackage[T1]{fontenc}
\usepackage[utf8]{inputenc}
\usepackage[american]{babel}
\usepackage{epsfig}
\usepackage{graphicx}
\usepackage{booktabs}
\usepackage{multirow}
\usepackage{dcolumn}
\usepackage{amsmath}
\usepackage{mathtools}
\usepackage{amsfonts}
\usepackage{amssymb}
\usepackage{epstopdf}
\usepackage{bm}
\usepackage{siunitx}
\usepackage{braket}
\usepackage{enumitem}
\usepackage{soul}
\usepackage[table]{xcolor}
\usepackage{color}
\usepackage{transparent}
\usepackage{pifont}

\definecolor{navyblue}{rgb}{0.0, 0.0, 0.5}
\definecolor{royalblue}{rgb}{0.25, 0.41, 0.88}
\definecolor{cadmiumgreen}{rgb}{0.0, 0.42, 0.24}
\definecolor{blue-violet}{rgb}{0.54, 0.17, 0.89}
\definecolor{darkviolet}{rgb}{0.58, 0.0, 0.83}
\definecolor{orange(colorwheel)}{rgb}{1.0, 0.5, 0.0}

\usepackage{hyperref}
\hypersetup{
    colorlinks=true, 
    linkcolor=royalblue, 
    citecolor=magenta}

\newcommand\ee{\end{equation}}
\newcommand\be{\begin{equation}}
\newcommand\eea{\end{eqnarray}}
\newcommand\bea{\begin{eqnarray}}

\newcommand{\om}{\Omega_m}
\newcommand{\omk}{\Omega_k}
\newcommand{\omrad}{\Omega_r}

\usepackage{booktabs}
\usepackage{multirow}
\usepackage{dcolumn}
\usepackage{colortbl}

\definecolor{magenta(process)}{rgb}{1.0, 0.0, 0.56}

\definecolor{darkspringgreen}{rgb}{0.09, 0.45, 0.27}

\definecolor{royalblue(web)}{rgb}{0.25, 0.41, 0.88}

\begin{document}

\title{Touch of Neutrinos on the Vacuum Metamorphosis: is the $H_0$ Solution Back?}

\author{Eleonora Di Valentino}
\email{eleonora.di-valentino@durham.ac.uk}
\affiliation{Institute for Particle Physics Phenomenology, Department of Physics, Durham University, Durham DH1 3LE, UK}

\author{Supriya Pan}
\email{supriya.maths@presiuniv.ac.in}
\affiliation{Department of Mathematics, Presidency University, 86/1 College Street, Kolkata 700073, India}

\author{Weiqiang Yang}
\email{d11102004@163.com}
\affiliation{Department of Physics, Liaoning Normal University, Dalian, 116029, P. R. China}

\author{Luis A. Anchordoqui}
\email{luis.anchordoqui@gmail.com}
\affiliation{Department of Physics and Astronomy,  Lehman College, City University of
  New York, NY 10468, USA}
\affiliation{Department of Physics,
 Graduate Center, City University
  of New York,  NY 10016, USA}
\affiliation{Department of Astrophysics,
 American Museum of Natural History, NY
 10024, USA}

\date{\today}

\preprint{}
\begin{abstract}
\noindent With the entrance of cosmology in its new era of high precision experiments, low- and high-redshift observations set off tensions in the measurements of both the present-day expansion rate ($H_0$) and the clustering of matter ($S_8$). We provide a simultaneous explanation of these tensions using the Parker-Raval Vacuum Metamorphosis (VM) model with the neutrino sector extended beyond the three massless Standard Model flavours and the curvature of the universe considered as a model parameter. To estimate the effect on cosmological observables we implement various extensions of the VM model in the standard \texttt{CosmoMC} pipeline and establish which regions of parameter space are empirically viable to resolve the $H_0$ and $S_8$ tensions. We constrain the parameter space employing the following data sets:
{\it (i)}~the cosmic microwave temperature and polarization data from the Planck mission, {\it (ii)}~Baryon Acoustic Oscillations (BAO) measurements, and {\it (iii)}~the Pantheon sample of Supernovae type Ia. We find that the likelihood analyses of the physically motivated VM model, which has the same number of free parameters as in the spatially-flat $\Lambda$CDM model, always gives  $H_0$ in agreement with the local measurements (even when BAO or Pantheon data are included) at the price of much larger $\chi^2$ than $\Lambda$CDM. The inclusion of massive neutrinos and extra relativistic species quantified through two well known parameters $\sum m_{\nu}$ and $N_{\rm eff}$, does not modify this result, and in some cases improves the goodness of the fit. In particular, for the original VM+$\sum m_\nu$+$N_{\rm eff}$ and the Planck+BAO+Pantheon dataset combination, we find evidence for $\sum m_{\nu}=0.80^{+0.18}_{-0.22}~{\rm eV}$ at more than $3\sigma$, no indication for extra neutrino species, $H_0=71.0\pm1.2$~km/s/Mpc in agreement with local measurements, and $S_8=0.755\pm0.032$ that solves the tension with the weak lensing measurements.
\end{abstract}
\maketitle
\section{Introduction} \label{sec:intro} 

The interval between the end of the twentieth century and the beginning of the twenty-first century is the golden period in cosmology. Some pioneering discoveries in this period, such as the observation of late time cosmic acceleration, the measurement of neutrino oscillations, and the detection  gravitational waves abruptly changed the traditional concept of our universe and opened new windows in front of the scientific community. Mostly, the observational data has been the key ingredient for such great discoveries, and the cosmology, we are currently witnessing today, has become more informative and precise. The concept of dark energy is the most exotic introduction in this period which is truly needed to understand the late time accelerating expansion of the universe and this contributes around 68\% of the total energy budget of the universe. The need for some cosmological constant was revived to explain this dark energy fluid and the $\Lambda$-Cold Dark Matter ($\Lambda$CDM hereafter) cosmology was found to accurately fit all the available
observational datasets. The $\Lambda$CDM cosmology, however, carries with it new serious questions. 
Apart from the fundamental and the long standing cosmological constant issue that is still unaddressed, the tension in the Hubble constant, $H_0$ is one of the greatest issues at present time within this paradigm.  
The estimated value of $H_0$ from the early measurements by the Planck team (within the $\Lambda$CDM picture) and the local distance ladders (in a model independent approach) are differing at many standard deviations. For instance, the 
value of $H_0$ using the early time measurements by the Planck team gives $H_0=67.27\pm 0.60$ km/s/Mpc at 68\% CL for Planck TTTEEE + lowE~\cite{Aghanim:2018eyx} (within the minimal $\Lambda$-Cold Dark Matter paradigm), while the estimated values of $H_0$ using the local distance ladders in a model independent approach are $H_0=74.03 \pm 1.42$ km/s/Mpc at 68\% CL~\cite{Riess:2019cxk}, and recently, $H_0=73.2 \pm 1.3$ km/s/Mpc at 68\% CL~\cite{Riess:2020fzl}. This is really intriguing because the measurement of a key cosmological parameter cannot be much different from two separate  measurements unless there are some potential systematic errors associated with the measurements, and this does not seem to be the case. Moreover, the $H_0$ tension has proved exceedingly challenging to understand theoretically, without fine-tuning.

There is also evidence of a growing tension between the Planck-preferred value and the local determination of $\sigma_8$, which gauges the amplitude of mass density fluctuations when smoothed with a top-hat filter of radius $8 h^{-1}~{\rm Mpc}$, where $h$ is the dimensionless Hubble constant~\cite{DiValentino:2020vvd}. More concretely, it is the combination $S_8= \sigma_8  (\Omega_m/0.3)^{0.5}$ that is constrained by large-scale structure data, where $\Omega_m$ is the present day value of the nonrelativistic matter density parameter. On the assumption of $\Lambda$CDM the Planck Collaboration reported $S_8 = 0.830 \pm 0.013$~\cite{Aghanim:2018eyx}, which is in $3\sigma$ tension with the result reported by KiDS-1000: $S_8 =
0.766^{+0.020}_{-0.014}$~\cite{Asgari:2020wuj}. The tension becomes $3.4\sigma$ if we consider a combination of BOSS and KV450: $S_8 =
0.728 \pm 0.026$~\cite{Troster:2019ean}. However, some datasets point to higher values of $S_8$, e.g. KiDS-450+GAMA for which $S_8 =
0.800^{+0.029}_{-0.027}$~\cite{vanUitert:2017ieu} or HSC SSP finding $S_8
=0.804^{+0.032}_{-0.029} $~\cite{Hamana:2019etx}. All in all, it seems of current interest to explore how to extend the $\Lambda$CDM concordance model of cosmology.

Inspired by the $H_0$ tension, various alternative approaches either modifying the matter sector or the gravitational sector of the universe, have been explored in the literature. The list is heavy for different variants of the cosmological models, for instance, the early dark energy \cite{Poulin:2018cxd,Karwal:2016vyq,Sakstein:2019fmf,Agrawal:2019lmo,Niedermann:2019olb,Niedermann:2020dwg}, phantom dark energy \cite{Yang:2018qmz,DiValentino:2019dzu,DiValentino:2020naf,Yang:2021flj}, interacting dark energy \cite{Kumar:2016zpg,DiValentino:2019ffd,Kumar:2019wfs,Lucca:2020zjb,DiValentino:2020vnx,Yang:2019uog,DiValentino:2017iww,Yang:2020uga,Yang:2018euj,Anchordoqui:2019amx,Pan:2020bur,Pan:2019gop,Yang:2019uzo,Pan:2019jqh,Anchordoqui:2020sqo}, emergent dark energy~\cite{Li:2019yem,Pan:2019hac,Li:2020ybr,Hernandez-Almada:2020uyr,Benaoum:2020qsi,Yang:2021egn}, modified gravity~\cite{Nunes:2018xbm,Cai:2019bdh,DAgostino:2020dhv,Wang:2020zfv}, decaying dark matter~\cite{Berezhiani:2015yta,Benisty:2019pxb,Vattis:2019efj,Blinov:2020uvz,Anchordoqui:2020djl}, and some others (see Refs.~\cite{DiValentino:2020zio,DiValentino:2021izs} and references therein for a comprehensive discussion in this direction).  
Usually, in most of the cases the alleviation of the $H_0$ tension is realized through the introduction of extra free parameters due to which the goodness of the model in fitting the data is worsened compared to the $\Lambda$CDM. We recall that the $\Lambda$CDM shows an excellent fit to most of the observational probes. So, this naturally raises an additional question regarding the goodness of the alternative cosmological models to fit the observational data even if the $H_0$ tension is alleviated. Generally speaking, this problem can be minimized for models with the smallest number of free parameters beyond the 6-parameter based $\Lambda$CDM, and it would become awesome if a cosmological model having only six parameters can  
really solve the $H_0$ tension. This inspired some investigators to look for alternative cosmological models mimicking the $\Lambda$CDM model in the number of free parameters. 
The construction of a 6-parameter model is not so difficult if we adopt the phenomenological route, but a cosmological model with the same degrees of freedom as in  $\Lambda$CDM while originated from some solid theoretical ground demands justification.

With this line of thinking, unlike other cosmological models with same number of degrees of freedom as in $\Lambda$CDM, the Vacuum Metamorphosis (VM) model could be the one having a solid quantum gravitational origin,  featuring a phase transition in the nature of the vacuum~\cite{Parker:1999td,Parker:1999ac,Parker:1999fc,Parker:2000pr,Parker:2001ws,Parker:2002xa,Parker:2003as}. This model has been investigated elsewhere leading to a solution of the $H_0$ tension~\cite{DiValentino:2017rcr,DiValentino:2020kha}. Herein, we proceed towards a complete investigation of this interesting $H_0$ solution by considering the impact of the neutrino sector in the evolution of VM cosmology.

The Standard Model (SM) of electroweak interactions includes three neutrino fields ($\nu_e, \nu_\mu, \nu_\tau$), which are left-handed partners of the three families of charged leptons ($e,\mu,\tau$). Because SM neutrinos only interact via weak interactions the off duty right-handed fields are absent in the SM by construction, and thereby SM neutrinos are massless. However, the phenomenon of neutrino oscillations observed in astrophysical and laboratory data implies that neutrinos have a finite mass, albeit very small~\cite{GonzalezGarcia:2007ib}. The presence of extra (sterile) neutrino species to accommodate neutrino masses must play an important role in the dynamics of our universe, since these neutrinos would modify the radiation energy density and as a result the determination of the cosmological parameters can equally be modified~\cite{Lesgourgues:2006nd,LaVacca:2009yp,Anchordoqui:2011nh,Anchordoqui:2012qu,Birrell:2012gg,Gerbino:2016sgw,Lattanzi:2017ubx,Nunes:2017xon, Yang:2017amu,Feng:2017usu,Vagnozzi:2018jhn}. It is therefore expected that the addition of sterile neutrinos into the VM framework will allow us to understand the cosmological behaviour of the model in a more comprehensive way. What's more, further understanding of some cosmological parameters could help elucidate the origin of the $H_0$ tension -- one of the main foci of this article. The standard parameters quantifying the neutrino sector in the universe are the total neutrino mass scale, $\sum m_{\nu}$, and the effective number of neutrino species, $N_{\rm eff}$~\cite{Steigman:1977kc}. In fact, since both $\sum m_\nu$ and $N_{\rm eff}$ are model dependent, one could equally assess their bounds and compare them to the same in other models. This is another motivation of this article where along with the behaviour of the Hubble constant, we investigate the bounds on $\sum m_{\nu}$ and $N_{\rm eff}$. 
Finally, we consider a one step further generalization of the VM model by adopting the curvature of the universe as a free parameter. 

The paper is structured as follows. In Sec.~\ref{sec:vm} we briefly review the generalities of the VM model and present the basic equations. In Sec.~\ref{sec:data} we discuss the observational datasets and the methodology that we use to constrain the various extensions of the VM model. Section~\ref{sec:results} elaborately describes the results and analyses of the models. Finally, in Sec.~\ref{sec:concl} we close the present work with a short summary.

\section{Vacuum Metamorphosis Revisited} \label{sec:vm}

In the VM paradigm, the universe undergoes a rapid transition from a standard Friedman-Lema\^{i}tre-Robetson-Walker (FLRW) spacetime dominated by cold dark matter to one containing significant contributions of vacuum energy and pressure~\cite{Parker:1999td,Parker:1999ac,Parker:1999fc,Parker:2000pr,Parker:2001ws,Parker:2002xa,Parker:2003as}. The vacuum pressure and energy density are both regulated by quantum effects of an ultra-light, minimally-coupled
scalar field. The negative vacuum pressure is responsible for the observed acceleration of the late-time universe, where the Ricci scalar $R$ remains constant. De facto, the scalar curvature plays the role of an order parameter steering a gravitational phase transition. Actually, Einstein's equations produce a back reaction on the metric which prevents $R$ from dropping below its critical value, yielding a kind of gravitational Lenz's law that keeps $R = m^2$, where at a microscopic level $m$ is proportional to the mass of the free quantized scalar field~\cite{Parker:2002xa}. For redshifts beyond the phase transition, the vacuum stress energy is negligible because $R > m^2$. In the local universe, however, we can distinguish two different regimes: {\it (i)}~$R> m^2$ in the vicinity of galaxies today, and so we observe no vacuum energy nearby; {\it (ii)}~$R \to m^2$ on large scales and accelerates the cosmic expansion of space.

The spacetime geometry is well-described by the maximally-symmetric FLRW line element
\begin{equation}
ds^2 = dt^2 - a^2(t) \left[\frac{dr^2 }{1 -k r^2}+ r^2 \ (d \theta^2 +
  \sin^2 \phi \ d\phi^2)\right] \,,   
\label{RW}
\end{equation}
where $(t,r,\theta,\phi$) are the comoving coordinates, $a(t)$ is the cosmic scale factor, and $k \ (= -1,0,1)$ parametrizes the curvature of homogeneous and isotropic spatial sections~\cite{Kolb:1990vq}. By matching a matter dominated scale parameter $a(t)$ and its first and second derivatives to the scale factor $a(t)$ of a constant $R$ universe at the transition redshift $z_t$, we uniquely determine the scale parameter $a(t)$ for $z< z_t$.
For $z > z_t$, the evolution of the universe is driven by the Friedmann equation for the Hubble parameter $H$,
\begin{equation}
  H^2(a) = \frac{8 \pi G}{3} \left\{\sum_i \rho_i(a) \right\} - \frac{k}{a^2} \,,
\label{eq:Friedmann}  
\end{equation}
where $G$ is the gravitational constant and the sum runs over  the energy densities $\rho_i$ of the various components of the cosmic fluid: CDM ($c$), baryons ($b$), and radiation ($r$). Note that the density of the dark energy $\rho_{\rm de}$ has been set to zero. The phase transition criticality condition is found to be 
\be 
R=6(\dot H+2H^2+ka^{-2})=m^2 \ , 
\label{criticality}
\ee
where dot denotes the derivative with respect to cosmic time.\footnote{The functional form of the criticality condition could be modified by adopting the Ricci invariant $R_{\mu \nu} R^{\mu \nu}$, Riemmann invariant $R_{\mu \nu \rho \sigma} R^{\mu \nu \rho \sigma}$, or Gauss-Bonnet invariant $Q = R^2 - 4 R_{\mu \nu} R^{\mu \nu} + R_{\mu \nu \rho \sigma} R^{\mu \nu \rho \sigma}$ as the order parameter in place of $R$~\cite{Caldwell:2005xb}. Throughout this paper we only consider variants of the standard VM model with criticality condition given by (\ref{criticality}).} Note that the Ricci scalar is a function of a single parameter, $m$, and so the scale factor for $z <z_t$ is fully determined by $m$. The values of $m$ and $z_t$ can thereby be expressed in terms of present-day observables:  
\be 
z_t=-1+\frac{3\Omega_m}{4(1-M-\omk-\omrad)} \ , \label{eq:zt}
\ee
with
\be 
\Omega_m=\frac{4}{3}\left[3M(1-M-\omk-\omrad)^3\right]^{1/4} \ , 
\label{eq:vmom}
\ee 
where $\Omega_m$ and $\omrad$ are respectively the densities (relative to the closure density) of matter and radiation, $\omk=-k/H_0^2$
is the curvature parameter, and $M=m^2/(12H_0^2)$. The expansion rate above and below 
the phase transition is described by 
\begin{widetext}
\bea 
H^2/H_0^2&=&\Omega_m (1+z)^3+\Omega_r (1+z)^4 +\omk (1+z)^2 
+M\left\{1-\left[3\left(\frac{4}{3\Omega_m}\right)^4 M(1-M-\omk-\omrad)^3\right]^{-1}
\right\}, \ z>z_t \label{eq:habove} \\
H^2/H_0^2&=&(1-M-\omk)(1+z)^4+\omk(1+z)^2+M\, ,\quad z\le z_t \, . \label{eq:hbelow} 
\eea
The equation of state parameter of the effective dark energy accelerating the present-day cosmic expansion is found to be 
\be 
w(z)=-1-\frac{1}{3}\frac{3\Omega_m (1+z)^{3}-4(1-M-\omk-\omrad)(1+z)^{4}}{M+(1-M-\omk-\omrad)(1+z)^{4}-\Omega_m (1+z)^{3}} \ , 
\ee 
\end{widetext}
whereas $w (z) = 0$ for $z > z_t$. The cosmic acceleration is driven by a phantom (i.e. $w < -1$) dark energy component~\cite{Caldwell:1999ew}, which asymptotically approaches a de Sitter phase (i.e. $w = -1$).

In (\ref{eq:Friedmann}) we have assumed that $\rho_{\rm de} = p_{\rm de} = 0$ to avoid introducing more than one parameter in the description of the gravitational phase transition.  We can now drop this supposition and in the spirit of~\cite{DiValentino:2020kha,DiValentino:2017rcr} extend the model assuming the massive scalar field has a vacuum expectation value (VEV), which manifests as a cosmological constant at high redshift. A point worth noting at this juncture is that under such assumption the cosmological model at high redshift ($z > z_t$) is purely $\Lambda$CDM,  while it is not at low redshift ($z < z_t$). The VEV, which is the magnitude of the high redshift cosmological constant, is a free parameter of this extended model. Hereafter, we denote the models as: original VM (if $w = 0$ for $z > z_t$) and VM-VEV (if $w=-1$ for $z> z_t$). In the VM-VEV model Eq.~(\ref{eq:vmom}) no longer describes the behavior of $\Omega_m$. For the VM-VEV model, we need to impose two extra conditions
$z_t\ge0$ and $\Omega_{\rm de} (z>z_t)\ge0$, which translate in a lower and upper bound on $\om$:
\be 
\frac{4}{3}(1-M-\omk-\omrad) \le\om\le \frac{4}{3}\left[3M(1-M-\omk-\omrad)^3\right]^{1/4}\, ; \label{eq:ombounds} 
\ee
for details see~\cite{DiValentino:2020kha,DiValentino:2017rcr}. In what follows we investigate the constraints of the data on both the original VM and VM-VEV models.

\section{Observational datasets} \label{sec:data}

In order to constrain the underlying cosmological scenarios, we have used various observational datasets. In the following we provide a succinct description of these data.

\begin{itemize}

\item {\bf CMB}: We consider the cosmic microwave background (CMB) temperature and polarization power spectra from the final release of Planck 2018 {\it plikTTTEEE+lowl+lowE}~\cite{Aghanim:2018eyx,Aghanim:2019ame}.

\item {\bf CMB lensing}: We consider the CMB lensing reconstruction power spectrum data obtained with a CMB trispectrum analysis~\cite{Aghanim:2018oex}.

\item {\bf BAO}: We have also considered Baryon Acoustic Oscillation (BAO) distance measurements from various astronomical missions such as 6dFGS~\cite{Beutler:2011hx}, SDSS MGS~\cite{Ross:2014qpa}, and BOSS DR12~\cite{Alam:2016hwk} as used by the Planck collaboration \cite{Aghanim:2018eyx}.

\item {\bf Pantheon}: Pantheon sample~\cite{Scolnic:2017caz} of the Type Ia Supernovae consisting of 1048 data points are also considered in the analysis.

\item {\bf R19}: Finally, we have considered the measurement of the Hubble constant provided by the SH0ES collaboration in a model independent approach~\cite{Riess:2019cxk}. The Hubble constant value is $H_0=74.03\pm1.42$ km/s/Mpc at 68\% CL and differs significantly from the Planck's estimation (assuming the $\Lambda$CDM background)~\cite{Aghanim:2018eyx}. We do not expect a significant variation in the measured value of $H_0$ when the luminosity distance is modified~\cite{Dhawan:2020xmp} to accommodate  $ 0< |\Omega_k| \ll 1$~\cite{Bernal:2016gxb}.

\end{itemize}
The baseline of the vacuum metamorphosis model consists of six parameters, namely, 
$\Omega_bh^2$ (baryon energy density), $\theta_{MC}$ (the ratio of sound horizon at decoupling to the angular diameter distance to last scattering), $\tau$ (the optical depth to reionization), the amplitude of the primordial scalar perturbations ($A_s$) and their spectral index ($n_s$), and $M$ (the 
vacuum metamorphosis parameter $M$ defined in Sec.~\ref{sec:vm}). We note that the parameter $M$ is related to the matter density $\Omega_m$ through Eq.~(\ref{eq:vmom}) in the original VM case, while it is a free parameter in the VM-VEV scenario. We then consider various extensions of this six parameter space model by including neutrinos and also the curvature of our universe. As noted in the Introduction, we describe the neutrino sector using the sum of three active neutrino masses $\sum m_\nu$ and the effective number of neutrino species $N_{\rm eff}$. The latter can be viewed as a convenient parametrization of the relativistic energy density of the Universe beyond that of photons, in units of the density of a single Weyl neutrino in the instantaneous decoupling limit. Therefore, we consider the following models assuming the spatially flat and non flat background: $\mathcal{X} +\sum m_{\nu}$, $\mathcal{X}+N_{\rm eff}$, $\mathcal{X}+\sum m_{\nu}+N_{\rm eff}$,  $\mathcal{X}+\sum m_{\nu}+N_{\rm eff}+\Omega_k$, for both the cases, i.e. the original $\mathcal{X} \equiv$ VM and the extended $\mathcal{X} \equiv$ VM-VEV.
We use in the analysis the flat uniform priors on the parameters reported in Table~\ref{tab:priors}.

Now, finally, to constrain all the scenarios described above, we have modified the Monte-Carlo Markov Chain code \texttt{CosmoMC}~\cite{Lewis:2002ah} a publicly free cosmological package (available from \url{http://cosmologist.info/cosmomc/}). The package supports the Planck 2018 likelihood~\cite{Aghanim:2019ame} having a precise convergence diagnostic based on the Gelman and Rubin statistics~\cite{Gelman:1992zz}. Additionally, this package appliances an efficient sampling of the posterior distribution which uses the fast/slow parameter decorrelations~\cite{Lewis:2013hha}.

\begin{table}
\begin{center}
\begin{tabular}{c|c}
\hline \hline
~~~~~~~~~Parameter~~~~~~~~~                    & ~~~~~~~~~Prior~~~~~~~~~\\
\hline 
$\Omega_{b} h^2$             & $[0.005,0.1]$\\
$\Omega_{c} h^2$             & $[0.001,0.99]$\\
$\tau$                       & $[0.01,0.8]$\\
$n_s$                        & $[0.8,1.2]$\\
${\rm{ln}}(10^{10}A_s)$         & $[1.6,3.9]$\\
$100\theta_{MC}$             & $[0.5,10]$\\ 
$M$                          & $[0.5,1]$\\ 
$\Omega_k$                   & $[-0.3,0.3]$\\
$\sum m_{\nu}$               & $[0.06,5]$\\
$N_{\rm eff}$                & $[0.05,10]$\\
\hline
\hline
\end{tabular}
\end{center}
\caption{Flat priors imposed on various free parameters of the underlying cosmological scenarios for the statistical analyses.}
\label{tab:priors}
\end{table}

\section{Cosmological Constraints}
\label{sec:results} 

We describe the observational constraints on various extensions of the VM model in a systematic way. Our baseline data is the CMB from the Planck 2018 release, and then we include other observational datasets in order to derive the constraints on the neutrino sector. For completeness, we have considered both spatially flat and non flat geometries of the Universe. In what follows we describe the observational constraints of the cosmological scenarios considered in this work, and we present a few selected cases in the triangular plots, to show the main correlations between the parameters.

Before proceeding, we pause to note one caveat of the VM and VM-VEV extensions discussed in this paper. Estimates of the sound horizon at the end of the baryonic-drag epoch, $r_{\rm drag}$, have been reported in~\cite{Arendse:2019hev}. These estimates are based on data from low-redshift probes and a set of polynomial parametrizations which are almost independent of the underlying cosmology. None of the VM nor VM-VEV extensions can accommodate the  $r_{\rm drag}$ estimates of~\cite{Arendse:2019hev} at the $1\sigma$ level.

\begin{table*}[tb]
\caption{68\% CL constraints and 95\% CL upper limits on the cosmological parameters of the original VM+$\sum m_{\nu}$ scenario using different dataset combinations explored in this work. The $\Delta \chi^2_{\rm bf}$ (best fit) is relative to the corresponding data best fits within $\Lambda$CDM.} 
\label{tab:mnu} 
\begin{center}
\resizebox{\textwidth}{!}{  
\begin{tabular}{ c |c c c c c c c} 
  \hline
 \hline
 Parameters & CMB & CMB+lensing & CMB+BAO & CMB+Pantheon & CMB+R19  & CMB+BAO+Pantheon & CMB+BAO+R19 \\ 

 \hline
  $\Omega_b h^2$ & $0.02232\pm0.00015$ & $0.02232^{+0.00016}_{-0.00014}$ & $0.02219\pm0.00012$ & $0.02181\pm0.00012$ & $0.02219\pm0.00012$ & $0.02213\pm0.00012$ & $0.02219\pm0.00012$ \\
  $100\theta_{MC}$ & $1.04080^{+0.00032}_{-0.00029}$ & $1.04081\pm0.00032$ & $1.04056\pm0.00030$ & $1.04004\pm0.00029$ & $1.04058\pm0.00029$ & $1.04051\pm0.00028$ & $1.04056\pm0.00028$ \\
  $\tau$ & $0.0534\pm0.0078$ & $0.0513^{+0.0082}_{-0.0073}$ & $0.0517\pm0.0079$ & $0.0480\pm0.0071$ & $0.0510\pm0.0076$ & $0.0535^{+0.0070}_{-0.0080}$ & $0.0515\pm0.0078$ \\
  $M$ & $0.9304^{+0.0093}_{-0.0051}$ & $0.931^{+0.011}_{-0.005}$ & $0.9159\pm0.0030$ & $0.844^{+0.013}_{-0.011}$ & $0.9173^{+0.0054}_{-0.0045}$ & $0.9081^{+0.0037}_{-0.0032}$ & $0.9156\pm0.0027$ \\
  ${\rm{ln}}(10^{10}A_s)$ & $3.044\pm0.016$ & $3.038\pm0.016$ & $3.043\pm0.016$ & $3.039\pm0.015$ & $3.042\pm0.016$ & $3.044^{+0.015}_{-0.017}$ & $3.043\pm0.016$ \\
  $n_s$ & $0.9631\pm0.0041$ & $0.9642^{+0.0044}_{-0.0039}$ & $0.9594\pm0.0033$ & $0.9467\pm0.0036$ & $0.9594\pm0.0033$ & $0.9588\pm0.0031$ & $0.9594\pm0.0032$ \\
  $\sum m_{\nu} [eV]$  & $<0.419$ & $<0.485$ & $0.38^{+0.12}_{-0.14}$ & $1.11^{+0.17}_{-0.20}$ & $0.33^{+0.11}_{-0.15}$ & $0.63^{+0.14}_{-0.17}$ & $0.39^{+0.13}_{-0.15}$ \\
   \hline
  $H_0 {\rm[km/s/Mpc]}$ & $79.1^{+3.0}_{-2.3}$ & $79.4^{+3.8}_{-2.4}$ & $74.44\pm0.78$ & $61.8\pm1.4$ & $74.8\pm1.4$ & $72.57\pm0.79$ & $74.34\pm0.69$ \\
  $\sigma_8$ & $0.914^{+0.036}_{-0.012}$ & $0.900^{+0.041}_{-0.018}$ & $0.863\pm0.034$ & $0.702\pm0.033$ & $0.876^{+0.038}_{-0.032}$ & $0.801\pm0.038$ & $0.859\pm0.036$ \\
  $S_8$ & $0.802\pm0.023$ & $0.788\pm0.015$ & $0.809\pm0.025$ & $0.813\pm0.028$ & $0.816^{+0.026}_{-0.023}$ & $0.777\pm0.029$ & $0.807\pm0.027$ \\
  $\Omega_m$ & $0.232^{+0.011}_{-0.021}$ & $0.231^{+0.012}_{-0.026}$ & $0.2641\pm0.0072$ & $0.404\pm0.022$ & $0.261^{+0.010}_{-0.012}$ & $0.2829^{+0.0077}_{-0.0089}$ & $0.2650\pm0.0066$ \\
  $r_{\rm drag}$ [Mpc] & $146.91\pm0.28$ & $147.01^{+0.28}_{-0.25}$ & $146.64\pm0.19$ & $145.74\pm0.20$ & $146.63\pm0.21$ & $143.97\pm0.15$ & $146.65\pm0.20$ \\
  \hline
    $\chi^2_{\rm bf}$ & $2768.85$ & $2779.008$ & $2803.17$ & $3831.966$ & $2772.228$ & $3894.196$ & $2803.11$ \\
$\Delta \chi^2_{\rm bf}$ & $-3.80$ & $-3.03$ & $+23.46$ & $+24.47$ & $-19.61$ & $+80.02$ & $+6.06$\\
   
 \hline
  \hline
\end{tabular}
}
\end{center}
\label{table}
\end{table*}

\subsection{Original VM}
In this section we will present the results obtained for extensions of the original VM scenario.

\subsubsection{VM+$\sum m_{\nu}$}

We first investigate a simple extension of the VM model considering the total neutrino mass $\sum m_{\nu}$ along with the original 6-parameters of the model. Thus, the free parameters of this scenario are seven. We explore several combinations of the cosmological probes, and we show the results in Table~\ref{tab:mnu}. 

We start investigating the constraints from CMB data alone, which are shown in the second column of Table~\ref{tab:mnu}. We first observe that $H_0$ takes a significantly larger value $H_0 = 79.1^{+3.0}_{-2.3}$ km/s/Mpc (68\% CL) than the minimal $\Lambda$CDM model using the same Planck dataset, where $H_0 \sim 67.4 \pm 0.5$ km/s/Mpc (68\% CL), but $1\sigma$ lower than the original VM model without $\sum m_{\nu}$ free to vary using Planck data (Table II of~\cite{DiValentino:2020kha}). This estimation is also larger than the R19 value \cite{Riess:2019cxk} ($H_0=74.03\pm1.42$ km/s/Mpc at 68\% CL), but can solve the tension within $1.5\sigma$. The scenario also indicates a relaxed bound on the total neutrino mass ($\sum m_{\nu}<0.419$ eV, 95\% CL upper limit) than the one obtained in a $\Lambda$CDM+$\sum m_{\nu}$ model ($\sum m_{\nu} < 0.257$ eV, 95\% CL upper limit) using Planck data. The same observations can be applied to the CMB+lensing dataset combination. Since CMB and R19 are consistent on the Hubble constant estimate, we can combine them together, obtaining the agreement on $H_0$ at the price of a neutrino mass scale different from zero at 95\% CL.

A similar interesting result is given when considering CMB+BAO. In fact, for this dataset combination $H_0=74.44\pm0.78$ km/s/Mpc (at 68\% CL) is fully consistent with R19 within $1\sigma$. In this case, a total neutrino mass different from zero is preferred at more than 99\% CL ($\sum m_{\nu}=0.38^{+0.12}_{-0.14}$eV at 68\% CL), and both $\sigma_8$ and $S_8$ are lowered with respect to the original VM without $\sum m_{\nu}$ free to vary~\cite{DiValentino:2020kha}, improving the agreement with the weak lensing data. Therefore, the addition of R19 in this case, i.e. CMB+BAO+R19, reduces  considerably the error bars, leaving unaltered the same features: solution of the Hubble constant and total neutrino mass at more than 3 standard deviations.

A completely different result is instead observed for the CMB+Pantheon dataset combination, which prefers a much larger value for $\sum m_{\nu}$ and a much lower value for $H_0$, in disagreement with the other cases. In particular, we have $\sum m_{\nu}=1.11^{+0.17}_{-0.20}$eV at 68\% CL and $H_0=61.8\pm1.4$
km/s/Mpc at 68\% CL. However, the inclusion of the BAO data, i.e. the CMB+BAO+Pantheon case, provides a striking result: the solution of both the $H_0$ and $S_8$ tensions within $1\sigma$, and a total neutrino mass above $5\sigma$. In particular, we find the following constraints at 68\% CL on key parameters: $H_0=72.57\pm0.79$ km/s/Mpc, $S_8=0.777\pm0.029$, and $\sum m_{\nu}=0.63^{+0.14}_{-0.17}$eV.

\begin{table*}[tb]
\caption{ 68\% CL constraints and 95\% CL upper limits on the cosmological parameters of the scenario original VM+$N_{\rm eff}$ for  different dataset combinations explored in this work. The $\Delta \chi^2_{\rm bf}$ (best fit) is relative to the corresponding data best fits within $\Lambda$CDM.} 
\label{tab:nnu} 
\begin{center}
\resizebox{\textwidth}{!}{  
\begin{tabular}{ c |c c c c c c c} 
  \hline
 \hline
 Parameters & CMB & CMB+lensing & CMB+BAO & CMB+Pantheon & CMB+R19  & CMB+BAO+Pantheon & CMB+BAO+R19 \\ 

 \hline
  $\Omega_b h^2$ & $0.02251\pm0.00025$ & $0.02239\pm0.00023$ & $0.02256\pm0.00023$ & $0.02283\pm0.00024$ & $0.02255\pm0.00024$ & $0.02232\pm0.00022$ & $0.02244\pm0.00021$ \\
  $100\theta_{MC}$ & $1.04073\pm0.00039$ & $1.04101\pm0.00036$ & $1.04039\pm0.00031$ & $1.03958\pm0.00034$ & $1.04034\pm0.00034$ & $1.04043\pm0.00031$ & $1.04041\pm0.00029$ \\
  $\tau$ & $0.0529\pm0.0081$ & $0.0514\pm0.0078$ & $0.0496\pm0.0077$ & $0.039^{+0.011}_{-0.007}$ & $0.0486\pm0.0077$ & $0.0455^{+0.0083}_{-0.0069}$ & $0.0481\pm0.0075$ \\
  $M$ & $0.9336^{+0.0076}_{-0.0058}$ & $0.9411^{+0.0040}_{-0.0034}$ & $0.9222\pm0.0024$ & $0.879\pm0.010$ & $0.9190^{+0.0049}_{-0.0042}$ & $0.9173\pm0.0024$ & $0.9205\pm0.0021$ \\
  ${\rm{ln}}(10^{10}A_s)$ & $3.046\pm0.018$ & $3.035\pm0.016$ & $3.048\pm0.017$ & $3.045^{+0.024}_{-0.016}$ & $3.047\pm0.017$ & $3.038^{+0.018}_{-0.015}$ & $3.044\pm0.016$ \\
  $n_s$ & $0.9698\pm0.0093$ & $0.9649\pm0.0090$ & $0.9724\pm0.0083$ & $0.9842\pm0.0091$ & $0.9721\pm0.0089$ & $0.9633\pm0.0082$ & $0.9682\pm0.0075$ \\
  $N_{\rm eff}$  & $3.18\pm0.19$ & $3.02\pm0.16$ & $3.32\pm0.14$ & $3.79\pm0.19$ & $3.34\pm0.18$ & $3.18\pm0.14$ & $3.25\pm0.13$ \\
   \hline
  $H_0 {\rm[km/s/Mpc]}$ & $80.4\pm2.4$ & $83.0\pm1.6$ & $76.61\pm0.93$ & $67.4^{+1.6}_{-1.8}$ & $75.7\pm1.2$ & $74.80\pm0.86$ & $75.91\pm0.76$ \\
  $\sigma_8$ & $0.9477\pm0.0099$ & $0.9384\pm0.0075$ & $0.9525\pm0.0084$ & $0.949^{+0.012}_{-0.008}$ & $0.9525\pm0.0087$ & $0.9480^{+0.0088}_{-0.0077}$ & $0.9505\pm0.0080$ \\
  $S_8$ & $0.815\pm0.028$ & $0.781\pm0.016$ & $0.859\pm0.011$ & $0.973\pm0.023$ & $0.869\pm0.017$ & $0.875\pm0.011$ & $0.8648\pm0.0096$ \\
  $\Omega_m$ & $0.222^{+0.012}_{-0.014}$ & $0.2078^{+0.0072}_{-0.0084}$ & $0.2441\pm0.0057$ & $0.316\pm0.016$ & $0.2500^{+0.0076}_{-0.0085}$ & $0.2556\pm0.0056$ & $0.2484\pm0.0048$ \\
  $r_{\rm drag}$ [Mpc] & $145.8\pm1.8$ & $147.5\pm1.5$ & $144.2\pm1.1$ & $139.6\pm1.5$ & $143.9\pm1.5$ & $145.2\pm1.1$ & $144.7\pm1.0$ \\
\hline
    $\chi^2_{\rm bf}$ & $2769.384$ & $2777.736$ & $2803.466$ & $3861.32$ & $2774.95$ & $3906.844$ & $2806.202$ \\
$\Delta \chi^2_{\rm bf}$ & $-3.27$ & $-4.30$ & $+23.76$ & $+53.82$ & $-16.89$ & $+92.66$ & $+9.15$\\
   
 \hline
  \hline
\end{tabular}
}
\end{center}
\label{table}
\end{table*}

\subsubsection{VM+$N_{\rm eff}$}

Secondly, we examine another simple extension of the VM model considering the effective number of neutrino species $N_{\rm eff}$ free to vary along with the original 6-parameters of the model. The results for different observational datasets are shown in Table~\ref{tab:nnu}.

Duplicating the procedure adopted in the previous section, we start investigating the constraints from CMB data alone. These are shown in the second column of Table~\ref{tab:nnu}. We find that $H_0$ takes a very high value when compared to the value obtained in a $\Lambda$CDM model using Planck data, but slightly lower than in the original VM model if $N_{\rm eff}$ is not allowed to vary in the fit~\cite{DiValentino:2020kha}. Moreover, the Hubble constant is also larger than the local measurements at about $2.3\sigma$.
This scenario also yields a $1\sigma$ shift higher value of $N_{\rm eff} = 3.18\pm0.19$ (68\% CL) when compared to the result from Planck ($N_{\rm eff} = 2.92^{+0.36}_{-0.37}$ at 68\% CL) in a $\Lambda$CDM+$N_{\rm eff}$ model. On the contrary, the CMB+lensing dataset combination gives $N_{\rm eff}$ almost identical to the standard value, so the constraints on the parameters are indistinguishable from those obtained in the original VM with $N_{\rm eff}$ fixed to the SM value 3.046~\cite{Mangano:2005cc}. In this scenario it is safe to combine CMB and R19 together, and the agreement on $H_0$ is obtained at the price of a $1\sigma$ indication for $\Delta N_{\rm eff} = N_{\rm eff} -3.046 >0$.

In addition, for this extension of the VM model, the results from the CMB+R19 dataset combination are very similar to those obtained when considering CMB+BAO. In fact, for the CMB+BAO dataset combination $H_0=76.61\pm0.93$ km/s/Mpc at 68\% CL, consistent with R19 at about $1.5\sigma$. In this case, a neutrino effective number different from the standard value is preferred at more than 95\% CL ($N_{\rm eff} = 3.32\pm0.14$ at 68\% CL). The addition of R19, i.e. CMB+BAO+R19,  decreases considerably the error bars on the $H_0$ determination, and also the indication for $\Delta N_{\rm eff} >0$ which is now just $1\sigma$.

An intriguing different result is instead observed for the CMB+Pantheon dataset combination, which prefers a much larger value for $N_{\rm eff}$ and $S_8$, and a much lower for $H_0$, in strong disagreement with predictions from the other dataset combinations. In particular, we have: $N_{\rm eff} = 3.79\pm0.19$ at 68\% CL, in disagreement at $3.9\sigma$ with the standard value and in agreement with $N_{\rm eff} =4$, and $H_0=67.4^{+1.6}_{-1.8}$ km/s/Mpc at 68\% CL, in disagreement at $3.1\sigma$ with R19. Finally, the inclusion of the BAO data to this combination, i.e. the CMB+BAO+Pantheon case, provides a solution of the $H_0$ tension within $1\sigma$, and $N_{\rm eff}$ is in agreement with the SM value of 3.046. We find the following constraints at 68\% CL on key parameters: $H_0=74.80\pm0.86$ km/s/Mpc and $N_{\rm eff} = 3.18\pm0.14$.


\begin{figure*}
\includegraphics[width=0.72\textwidth]{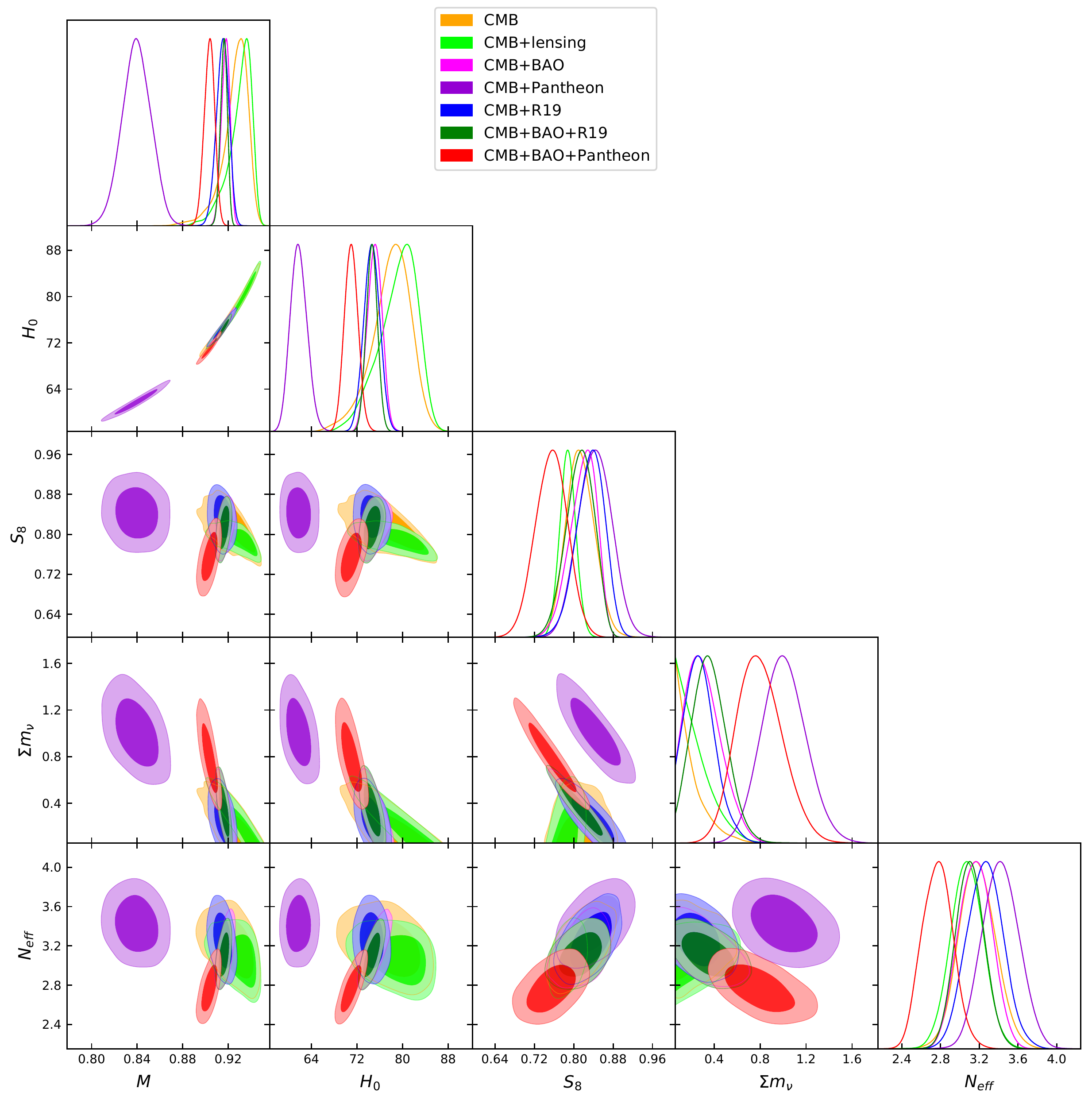}
\caption{68\% and 95\% CL constraints on the original VM+$N_{\rm eff}$+$\sum m_{\nu}$ case. }
\label{fig:mnunnu}
\end{figure*}


\begin{table*}[tb]
\caption{ 68\% CL constraints and 95\% CL upper limits on the cosmological parameters of the scenario original VM+$N_{\rm eff}$+$\sum m_{\nu}$ for  different dataset combinations explored in this work. The $\Delta \chi^2_{\rm bf}$ (best fit) is relative to the corresponding data best fits within $\Lambda$CDM.
} 
\label{tab:mnunnu} 
\begin{center}
\resizebox{\textwidth}{!}{  
\begin{tabular}{ c |c c c c c c c} 
  \hline
 \hline
 Parameters & CMB & CMB+lensing & CMB+BAO & CMB+Pantheon & CMB+R19  & CMB+BAO+Pantheon & CMB+BAO+R19 \\ 

 \hline
  $\Omega_b h^2$ & $0.02245\pm0.00025$ & $0.02236\pm0.00023$ & $0.02234\pm0.00026$ & $0.02225\pm0.00026$ & $0.02245\pm0.00025$ & $0.02174\pm0.00026$ & $0.02226\pm0.00013$ \\
  $100\theta_{MC}$ & $1.04060\pm0.00041$ & $1.04079\pm0.00039$ & $1.04049\pm0.00032$ & $1.03971\pm0.00034$ & $1.04037\pm0.00034$ & $1.04075\pm0.00033$ & $1.04051\pm0.00031$ \\
  $\tau$ & $0.0530\pm0.0077$ & $0.0519\pm0.0075$ & $0.0520\pm0.0077$ & $0.0498\pm0.0075$ & $0.0513\pm0.0077$ & $0.0518\pm0.0074$ & $0.0517\pm0.0078$ \\
  $M$ & $0.926^{+0.013}_{-0.006}$ & $0.930^{+0.013}_{-0.005}$ & $0.9176^{+0.0041}_{-0.0034}$ & $0.839\pm0.013$ & $0.9150^{+0.0058}_{-0.0050}$ & $0.9034^{+0.0048}_{-0.0043}$ & $0.9161^{+0.0032}_{-0.0028}$ \\
  ${\rm{ln}}(10^{10}A_s)$ & $3.047\pm0.017$ & $3.040\pm0.016$ & $3.046\pm0.017$ & $3.053\pm0.017$ & $3.048\pm0.017$ & $3.032\pm0.017$ & $3.044\pm0.017$ \\
  $n_s$ & $0.9686\pm0.0089$ & $0.9653\pm0.0086$ & $0.9652\pm0.0096$ & $0.9651\pm0.0099$ & $0.9692\pm0.0093$ & $0.9429\pm0.0097$ & $0.9622\pm0.0084$ \\
  $\sum m_{\nu} [eV]$  & $<0.468$ & $<0.520$ & $0.31^{+0.10}_{-0.19}$ & $1.01\pm0.19$ & $0.28^{+0.10}_{-0.16}$ & $0.80^{+0.18}_{-0.22}$ & $0.36^{+0.13}_{-0.17}$ \\
  $N_{\rm eff}$  & $3.18\pm0.19$ & $3.07\pm0.17$ & $3.16\pm0.17$ & $3.42\pm0.19$ & $3.26\pm0.19$ & $2.77\pm0.16$ & $3.11\pm0.15$ \\
   \hline
   \hline
  $H_0 {\rm[km/s/Mpc]}$ & $78.0^{+3.8}_{-2.6}$ & $79.2^{+4.1}_{-2.4}$ & $75.1\pm1.3$ & $61.8^{+1.4}_{-1.6}$ & $74.6\pm1.4$ & $71.0\pm1.2$ & $74.60\pm0.97$ \\
  $\sigma_8$ & $0.913^{+0.040}_{-0.014}$ & $0.898^{+0.043}_{-0.018}$ & $0.883^{+0.051}_{-0.034}$ & $0.729\pm0.037$ & $0.893^{+0.042}_{-0.030}$ & $0.757\pm0.045$ & $0.869^{+0.044}_{-0.038}$ \\
  $S_8$ & $0.813^{+0.027}_{-0.030}$ & $0.788\pm0.016$ & $0.819^{+0.032}_{-0.022}$ & $0.843\pm0.034$ & $0.834^{+0.031}_{-0.025}$ & $0.755\pm0.032$ & $0.813^{+0.031}_{-0.026}$ \\
  $\Omega_m$ & $0.239^{+0.013}_{-0.027}$ & $0.233^{+0.012}_{-0.028}$ & $0.258^{+0.009}_{-0.012}$ & $0.402\pm0.022$ & $0.262^{+0.010}_{-0.012}$ & $0.299\pm0.013$ & $0.2627^{+0.0081}_{-0.0092}$ \\
  $r_{\rm drag}$ [Mpc]& $145.7\pm1.8$ & $146.8\pm1.5$ & $145.7\pm 1.5$ & $142.6\pm1.5$ & $144.7\pm1.6$ & $149.1\pm1.4$ & $146.1\pm1.3$ \\
  \hline
  $\chi^2_{\rm bf}$ & $2769.976$ & $2778.582$ & $2803.986$ & $3828.246$ & $2773.55$ & $3891.388$ & $2802.988$ \\
  $\Delta \chi^2_{\rm bf}$ & $-2.67$ & $-3.46$ & $+24.28$ & $+20.75$ & $-18.29$ & $+77.21$ & $+5.94$\\
   
 \hline
  \hline
\end{tabular}
}
\end{center}
\label{table}
\end{table*}

\subsubsection{VM+$N_{\rm eff}$+$\sum m_{\nu}$}

In this section we study the original VM model along with the total neutrino mass $\sum m_{\nu}$ and the effective number of neutrino species $N_{\rm eff}$ varying as free parameters of the model at the same time. The results for different observational datasets are given in Table~\ref{tab:mnunnu}, and the 1D posterior distributions and the 2D contour plots are shown in Fig.~\ref{fig:mnunnu}.

Considering the CMB data alone (second column of Table~\ref{tab:mnunnu}) we can see that $H_0 $ lowers significantly with respect to the previous cases. In particular now $H_0 = 78.0^{+3.8}_{-2.6}$ km/s/Mpc (68\% CL), helping in solving the tension with R19 at $1.3\sigma$. This scenario also indicates a slightly relaxed bound on the total neutrino mass ($\sum m_{\nu}<0.468$~eV, 95\% CL upper limit), and an unaltered constraint on $N_{\rm eff}=3.18\pm0.19$ at 68\% CL. In fact, these two parameters do not show a significant correlation in Fig.~\ref{fig:mnunnu}. 
The same observations can be applied to the CMB+lensing dataset combination. 
Because of the agreement of the CMB and R19, we can combine them together, obtaining the agreement on $H_0$ at the price of both a total neutrino mass and a $\Delta N_{\rm eff}$ different from zero at 68\% CL. 

As happened with the previous VM extensions, the CMB+R19 dataset gives an interesting result, which is similar to the one obtained for the CMB+BAO combination. In fact, for CMB+BAO we have $H_0=75.1\pm1.3$ km/s/Mpc at 68\% CL, fully consistent with R19 within $1\sigma$. In addition, for the CMB+BAO datasets, we have a total neutrino mass different from zero at more than 68\% CL ($\sum m_{\nu}=0.31^{+0.10}_{-0.19}$eV at 68\% CL), and an effective number of neutrino species in agreement with the SM value. The addition of R19, i.e. the CMB+BAO+R19 combination, improves considerably the constraints, with preference for a total neutrino mass at $2\sigma$ and $\Delta N_{\rm eff}=0$.

Finally, also in this extended scenario, the CMB+Pantheon dataset is in disagreement with the other dataset combinations, as it prefers a much larger value for both $\sum m_{\nu}$ and $\Delta N_{\rm eff}>0$ at more than 95\% CL, and a much lower value for $H_0$. This is possible because the well-known strong correlation between $N_{\rm eff}$ and $H_0$ is absent in this extended VM model (see Fig.~\ref{fig:mnunnu}).
In particular, we have at 68\% CL $\sum m_{\nu}=1.01\pm0.19$eV, $N_{\rm eff}=3.42\pm0.19$, and $H_0=61.8^{+1.4}_{-1.6}$ km/s/Mpc. For this VM extension, the inclusion of the BAO data in the CMB+BAO+Pantheon combination changes completely all the constraints. Both the $H_0$ and $S_8$ tensions are solved within $1\sigma$, a total neutrino mass above 99\% CL appears, and the effective number of equivalent neutrinos shifts towards lower values (more than $1\sigma$ below the expected value). In particular, we find the following constraints at 68\% CL on key parameters: $H_0=71.0\pm1.2$ km/s/Mpc, $S_8=0.755\pm0.032$, $\sum m_{\nu}=0.80^{+0.18}_{-0.22}$eV, and $N_{\rm eff}=2.77\pm0.16$.


\begin{figure*}
\includegraphics[width=0.72\textwidth]{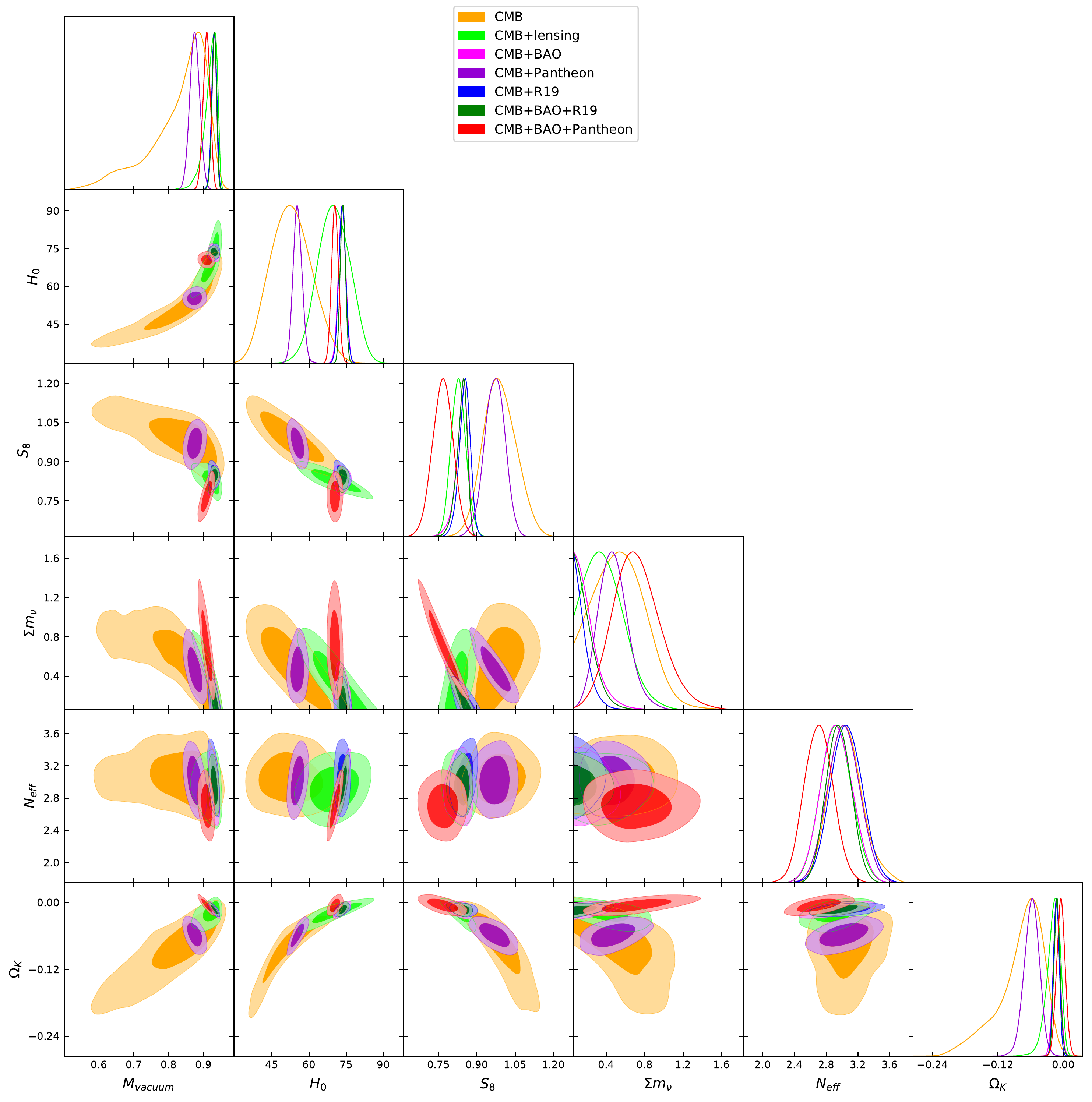}
\caption{68\% and 95\% CL constraints on the original VM+$N_{\rm eff}$+$\sum m_{\nu}$+$\omk$ case. }
\label{fig:mnunnuomk}
\end{figure*}


\begin{table*}[tb]
\caption{ 68\% CL constraints and 95\% CL upper limits on the cosmological parameters of the scenario original VM+$N_{\rm eff}$+$\sum m_{\nu}$+$\omk$ for different dataset combinations explored in this work. The $\Delta \chi^2_{\rm bf}$ (best fit) is relative to the corresponding data best fits within $\Lambda$CDM.
} 
\label{tab:mnunnuomk} 
\begin{center}
\resizebox{\textwidth}{!}{  
\begin{tabular}{ c |c c c c c c c} 
  \hline
 \hline
 Parameters & CMB & CMB+lensing & CMB+BAO & CMB+Pantheon & CMB+R19  & CMB+BAO+Pantheon & CMB+BAO+R19 \\ 

 \hline
  $\Omega_b h^2$ & $0.02251\pm0.00027$ & $0.02230\pm0.00024$ & $0.02232\pm0.00025$ & $0.02249\pm0.00027$ & $0.02250\pm0.00025$ & $0.02174^{+0.00050}_{-0.00049}$ & $0.02236\pm0.00022$ \\
  $100\theta_{MC}$ & $1.04100\pm0.00047$ & $1.04108\pm0.00046$ & $1.04111\pm0.00045$ & $1.04098\pm0.00045$ & $1.04100\pm0.00044$ & $1.04095^{+0.00091}_{-0.00087}$ & $1.04107\pm0.00041$ \\
  $\tau$ & $0.0471\pm0.0088$ & $0.0485\pm0.0081$ & $0.0517\pm0.0078$ & $0.0472\pm0.0080$ & $0.0522\pm0.0078$ & $0.051\pm0.015$ & $0.0517\pm0.0078$ \\
  $M$ & $0.830^{+0.096}_{-0.034}$ & $0.919^{+0.023}_{-0.010}$ & $0.9300^{+0.0072}_{-0.0061}$ & $0.875\pm0.014$ & $0.9302^{+0.0078}_{-0.0065}$ & $0.908^{+0.019}_{-0.021}$ & $0.9301^{+0.0070}_{-0.0059}$ \\
  ${\rm{ln}}(10^{10}A_s)$ & $3.024\pm0.022$ & $3.023\pm0.019$ & $3.033\pm0.018$ & $3.023\pm0.020$ & $3.037\pm0.018$ & $3.027\pm0.035$ & $3.034\pm0.018$ \\
  $n_s$ & $0.968\pm0.010$ & $0.9618\pm0.0090$ & $0.9618\pm0.0091$ & $0.9675\pm0.0097$ & $0.9683\pm0.0092$ & $0.942\pm0.018$ & $0.9630\pm0.0078$ \\
  $\sum m_{\nu} [eV]$  & $0.54^{+0.24}_{-0.30}$ & $0.39^{+0.14}_{-0.26}$ & $<0.446$ & $0.47^{+0.14}_{-0.16}$ & $<0.323$ & $0.72^{+0.50}_{-0.47}$ & $<0.397$ \\
  $N_{\rm eff}$  & $3.05^{+0.18}_{-0.23}$ & $2.92\pm0.19$ & $2.93\pm0.20$ & $3.02\pm0.20$ & $3.05\pm0.20$ & $2.71^{+0.36}_{-0.34}$ & $2.96\pm0.16$ \\
  $\Omega_k$  & $-0.077^{+0.050}_{-0.022}$ & $-0.019^{+0.015}_{-0.009}$ & $-0.0124\pm0.0063$ & $-0.058^{+0.014}_{-0.013}$ & $-0.0130\pm0.0052$ & $-0.005\pm0.015$ & $-0.0117\pm0.0055$ \\
   \hline
   \hline
  $H_0 {\rm[km/s/Mpc]}$ & $53\pm8$ & $70.0\pm6.4$ & $73.4\pm1.5$ & $55.4\pm1.8$ & $73.4\pm1.4$ & $70.5\pm2.6$ & $73.7\pm1.0$ \\
  $\sigma_8$ & $0.737\pm0.085$ & $0.828^{+0.068}_{-0.054}$ & $0.887^{+0.038}_{-0.020}$ & $0.766\pm0.032$ & $0.902^{+0.027}_{-0.016}$ & $0.766^{+0.098}_{-0.094}$ & $0.892^{+0.032}_{-0.019}$ \\
  $S_8$ & $0.990^{+0.060}_{-0.067}$ & $0.828^{+0.027}_{-0.030}$ & $0.839^{+0.027}_{-0.015}$ & $0.970\pm0.039$ & $0.852^{+0.023}_{-0.019}$ & $0.767^{+0.077}_{-0.078}$ & $0.840^{+0.025}_{-0.015}$ \\
  $\Omega_m$ & $0.58^{+0.12}_{-0.23}$ & $0.308^{+0.041}_{-0.075}$ & $0.269^{+0.010}_{-0.012}$ & $0.482\pm0.031$ & $0.268^\pm0.011$ & $0.301^{+0.026}_{-0.025}$ & $0.2663\pm0.0077$ \\
  $r_{\rm drag}$ [Mpc]& $147.2^{+2.2}_{-1.9}$ & $148.4\pm1.9$ & $148.3\pm 2.0$ & $147.5\pm1.9$ & $147.2\pm1.9$ & $149.9^{+3.7}_{-3.6}$ & $148.0\pm1.5$ \\
 \hline
  $\chi^2_{\rm bf}$ & $2765.572$ & $2777.134$ & $2795.998$ & $3801.384$ & $2764.186$ & $3891.750$ & $2795.730$ \\
 $\Delta \chi^2_{\rm bf}$ & $-7.08$ & $-4.91$ & $+16.29$ & $-6.12$ & $-27.65$ & $+77.57$ & $-1.32$\\   
 \hline
  \hline
\end{tabular}
}
\end{center}
\label{table}
\end{table*}

\subsubsection{VM+$N_{\rm eff}$+$\sum m_{\nu}$+$\omk$}
\label{omk1}

For completeness, in this last section we consider an extension of the original VM model where, together with the 6-parameters of the original VM model, the total neutrino mass $\sum m_{\nu}$, the effective number of neutrinos $N_{\rm eff}$, and the curvature energy density $\omk$ are considered as free parameters. The results for the different observational dataset combinations are given in Table~\ref{tab:mnunnuomk}, and in Fig.~\ref{fig:mnunnuomk} it is shown a triangular plot with the key parameters.

Regarding the CMB data alone we can notice that $H_0 $ lowers significantly with respect to the previous cases, but with very large error bars. In particular we find $H_0 = 53\pm8$ km/s/Mpc (68\% CL), in tension with R19 at $2\sigma$. In this scenario, because of the $\sum m_{\nu}$-$\omk$ correlation (see Fig.~\ref{fig:mnunnuomk}) we have $1\sigma$ indication for a total neutrino mass ($\sum m_{\nu}=0.54^{+0.24}_{-0.30}$ eV at 68\% CL), $N_{\rm eff}=3.05^{+0.18}_{-0.23}$ at 68\% CL completely in agreement with the SM $3.046$, and $\omk=-0.077^{+0.050}_{-0.022}$ at 68\% CL, preferring a closed Universe at more than 95\% CL. 
Different, in this case, is the result obtained with the CMB+lensing dataset combination. In fact, the lensing dataset contributes to break the geometrical degeneracy of the parameters, giving $H_0$ in agreement with R19 within $1\sigma$ ($H_0 = 70.0\pm6.4$ km/s/Mpc at 68\% CL). However, even in this case, there is $1\sigma$ indication for a total neutrino mass, $N_{\rm eff}$ in agreement with the standard value, and the preference for a closed Universe at 68\% CL.

Since the CMB is not in strong tension with R19, we can combine them together, obtaining the agreement on $H_0$ at the price of a closed Universe at $3$ standard deviations ($\omk=-0.0130\pm0.0052$ at 68\% CL), while both a total neutrino mass and a $\Delta N_{\rm eff}$ agree with zero within the 68\% CL. 
Similar interesting results are obtained for the CMB+BAO combination. In fact, for CMB+BAO we have $H_0=73.4\pm1.5$ km/s/Mpc at 68\% CL, fully consistent with both R19 and CMB+R19, within $1$ standard deviation. Also in this case, we have both a total neutrino mass and a $\Delta N_{\rm eff}$ in agreement with zero within the 68\% CL, and a preference for a closed Universe at $2$ standard deviations ($\omk=-0.0124\pm0.0063$ at 68\% CL).  
The inclusion of the Hubble constant prior R19 for the CMB+BAO+R19 combination, gives the same result, but with smaller error bars.

In this VM extension, for CMB+Pantheon we have a complete agreement with the CMB alone data sample. In fact, because of the $\sum m_{\nu}$-$\omk$ correlation (see Fig.~\ref{fig:mnunnuomk}), we have evidence at more than $3\sigma$ for a total neutrino mass ($\sum m_{\nu}=0.47^{+0.14}_{-0.16}$ eV at 68\% CL), $N_{\rm eff}=3.02\pm0.20$ at 68\% CL completely consistent with the the SM value, and $\omk=-0.058^{+0.014}_{-0.013}$ at 68\% CL, preferring a closed Universe at more than 99\% CL. For the CMB+Pantheon dataset combination, the Hubble constant value is much lower than R19, several standard deviations away, and equal to $H_0=55.4\pm1.8$ km/s/Mpc at 68\% CL.
For this VM extension, the BAO data are in disagreement with CMB+Pantheon, therefore the CMB+BAO+Pantheon combination is not reliable.

\begin{table*}[tb]
\caption{ 68\% CL constraints and 95\% CL upper limits on the cosmological parameters of the scenario VM-VEV+$\sum m_{\nu}$ for different dataset combinations explored in this work. The $\Delta \chi^2_{\rm bf}$ (best fit) is relative to the corresponding data best fits within $\Lambda$CDM.
} 
\label{tab:Mmnu} 
\begin{center}
\resizebox{\textwidth}{!}{  
\begin{tabular}{ c |c c c c c c c} 
  \hline
 \hline
 Parameters & CMB & CMB+lensing & CMB+BAO & CMB+Pantheon & CMB+R19  & CMB+BAO+Pantheon & CMB+BAO+R19 \\ 

 \hline
  $\Omega_b h^2$ & $0.02235\pm0.00016$ & $0.02237\pm0.00015$ & $0.02228\pm0.00014$ & $0.02226\pm0.00016$ & $0.02232\pm0.00015$ & $0.02227\pm0.00014$ & $0.02230\pm0.00014$ \\
  $\Omega_c h^2$ & $0.1202\pm0.0014$ & $0.1200\pm0.0013$ & $0.1210\pm0.0012$ & $0.1213\pm0.0014$ & $0.1204\pm0.0014$ & $0.1213\pm0.0012$ & $0.1209\pm0.0012$ \\
  $100\theta_{MC}$ & $1.04084\pm0.00033$ & $1.04087\pm0.00032$ & $1.04074\pm0.00030$ & $1.04071\pm0.00033$ & $1.04080\pm0.00033$ & $1.04075\pm0.00029$ & $1.04077\pm0.00031$ \\
  $\tau$ & $0.0545\pm0.0080$ & $0.0538\pm0.0076$ & $0.0537\pm0.0077$ & $0.0536\pm0.0076$ & $0.0545\pm0.0075$ & $0.0539\pm0.0077$ & $0.0539\pm0.0076$ \\
  $M$ & $0.912^{+0.022}_{-0.009}$ & $0.918^{+0.017}_{-0.008}$ & $0.8954^{+0.0017}_{-0.0043}$ & $0.8924^{+0.0021}_{-0.0025}$ & $0.9040^{+0.0061}_{-0.0085}$ & $0.8922^{+0.0011}_{-0.0017}$ & $0.8958^{+0.0017}_{-0.0043}$ \\
  ${\rm{ln}}(10^{10}A_s)$ & $3.045\pm0.016$ & $3.043\pm0.015$ & $3.045\pm0.016$ & $3.045\pm0.016$ & $3.045\pm0.015$ & $3.046\pm0.016$ & $3.046\pm0.016$ \\
  $n_s$ & $0.9644\pm0.0044$ & $0.9649\pm0.0043$ & $0.9625\pm0.0040$ & $0.9616\pm0.0045$ & $0.9594\pm0.0033$ & $0.9618\pm0.0039$ & $0.9628\pm0.0041$ \\
  $\sum m_{\nu} [eV]$  & $<0.358$ & $<0.319$ & $<0.354$ & $<0.579$ & $<0.399$ & $<0.352$ & $<0.321$ \\
   \hline
  $H_0 {\rm[km/s/Mpc]}$ & $76.0^{+4.4}_{-2.3}$ & $77.1^{+2.9}_{-2.3}$ & $73.19\pm0.59$ & $72.5^{+1.2}_{-0.6}$ & $74.5^{+1.0}_{-1.2}$ & $72.73^{+0.54}_{-0.43}$ & $73.33^{+0.48}_{-0.55}$ \\
  $\sigma_8$ & $0.877\pm0.029$ & $0.884^{+0.027}_{-0.022}$ & $0.854^{+0.024}_{-0.016}$ & $0.840^{+0.037}_{-0.015}$ & $0.863^{+0.025}_{-0.016}$ & $0.851^{+0.024}_{-0.016}$ & $0.856^{+0.021}_{-0.015}$ \\
  $S_8$ & $0.800\pm0.018$ & $0.795\pm0.014$ & $0.812^{+0.019}_{-0.016}$ & $0.809^{+0.021}_{-0.017}$ & $0.804^{+0.019}_{-0.016}$ & $0.815^{+0.019}_{-0.016}$ & $0.812\pm0.017$ \\
  $\Omega_m$ & $0.250^{+0.017}_{-0.029}$ & $0.243^{+0.014}_{-0.020}$ & $0.2713^{+0.0054}_{-0.0065}$ & $0.278^{+0.007}_{-0.014}$ & $0.2606\pm0.0093$ & $0.2753^{+0.0052}_{-0.0065}$ & $0.2697\pm0.0054$ \\
  $r_{\rm drag}$ [Mpc] & $147.04\pm0.30$ & $147.09\pm0.28$ & $146.89\pm0.27$ & $146.83\pm0.30$ & $147.01\pm0.30$ & $146.83\pm0.26$ & $146.92\pm0.27$ \\
 \hline
  $\chi^2_{\rm bf}$ & $2767.962$ & $2778.142$ & $2791.026$ & $3840.314$ & $2772.594$ & $3854.916$ & $2792.516$ \\
 $\Delta \chi^2_{\rm bf}$ & $-4.69$ & $-3.90$ & $+11.32$ & $+32.81$ & $-19.25$ & $+40.74$ & $-4.53$\\  

 \hline
  \hline
\end{tabular}
}
\end{center}
\label{table}
\end{table*}
\subsection{VM-VEV}
In this section we will present the results obtained for extensions of the VM-VEV scenario.

\subsubsection{VM-VEV+$\sum m_{\nu}$}

We first consider the simple extension of the VM-VEV model where the total neutrino mass $\sum m_{\nu}$ has been considered free to vary along with 7-parameters of the VM-VEV model. We analysed several combinations of the cosmological probes listed in this paper, we show the results in Table~\ref{tab:Mmnu}.  As already observed in Ref.~\cite{DiValentino:2020kha}, in the VM-VEV scenario the additional degree of freedom lowers the Hubble constant value, improving the agreement with R19. Additionally, the well known $\sum m_{\nu}$-$H_0$ negative correlation lowers still more the $H_0$ value with respect to the case in which $\sum m_{\nu}$ is fixed to the arbitrary value of $0.06~{\rm eV}$ of the standard $\Lambda$CDM scenario.

If we look at the constraints for CMB alone (second column of Table~\ref{tab:Mmnu}), we see that now $H_0 = 76.0^{+4.4}_{-2.3}$ km/s/Mpc at 68\% CL, much larger than the estimate obtained by Planck within the minimal $\Lambda$CDM model, and in $1\sigma$ agreement with R19.
The CMB only case presents also a relaxed bound on the total neutrino mass ($\sum m_{\nu}<0.358$ eV, 95\% upper limit) than Planck in a $\Lambda$CDM+$\sum m_{\nu}$ model, but stronger than the original VM+$\sum m_{\nu}$ of Table~\ref{tab:mnu}). The very same cosmological parameters can be inferred from the CMB+lensing dataset combination. 
Combining CMB and R19 in this case does not give any indication for a total neutrino mass different from zero, and slightly relaxes its upper bound ($\sum m_{\nu}<0.399$ eV, 95\% upper limit).

In this minimal VM-VEV extension, the CMB+BAO dataset combination lowers significantly the central value of  Hubble constant and reduces drastically its error bars when compared to the CMB alone case. The CMB+BAO dataset leads to $H_0=73.19\pm0.59$ km/s/Mpc at 68\% CL, still fully consistent with R19 within $1\sigma$. In this case, we have almost the same upper limit of the CMB alone case: $\sum m_{\nu}<0.354$eV at 95\% CL. If to this dataset combination we now add R19, the fit favors a further reduction of the $H_0$ error bars and strengthens the total neutrino mass upper limit.

Contrary to what happens in the original VM extension, the CMB+Pantheon dataset combination is now completely in agreement with the other cases. In particular, we have $\sum m_{\nu}<0.579$eV at 95\% CL, slightly relaxed with respect to the CMB only case, and $H_0=72.5^{+1.2}_{0.6}$ km/s/Mpc at 68\% CL, now perfectly consistent with R19. The inclusion of the BAO data for the CMB+BAO+Pantheon case, confirms completely these findings, solving $H_0$ and making stronger the upper $\sum m_{\nu}$ limit. In particular we find $H_0=72.73^{+0.54}_{-0.43}$ km/s/Mpc at 68\% CL and $\sum m_{\nu}<0.352$eV at 95\% CL.

\begin{table*}[tb]
\caption{ 68\% CL constraints and 95\% CL upper limits on the cosmological parameters of the scenario VM-VEV+$N_{\rm eff}$ for different dataset combinations explored in this work. The $\Delta \chi^2_{\rm bf}$ (best fit) is relative to the corresponding data best fits within $\Lambda$CDM.
} 
\label{tab:Mnnu} 
\begin{center}
\resizebox{\textwidth}{!}{  
\begin{tabular}{ c |c c c c c c c} 
  \hline
 \hline
 Parameters & CMB & CMB+lensing & CMB+BAO & CMB+Pantheon & CMB+R19  & CMB+BAO+Pantheon & CMB+BAO+R19 \\ 

 \hline
  $\Omega_b h^2$ & $0.02228\pm0.00022$ & $0.02229\pm0.00021$ & $0.02223\pm0.00021$ & $0.02239\pm0.00022$ & $0.02255\pm0.00024$ & $0.02232\pm0.00022$ & $0.02244\pm0.00021$ \\
  $\Omega_c h^2$ & $0.1181\pm0.0030$ & $0.1175\pm0.0028$ & $0.1201\pm0.0029$ & $0.1219\pm0.0030$ & $0.1204\pm0.0014$ & $0.1213\pm0.0012$ & $0.1209\pm0.0012$ \\
  $100\theta_{MC}$ & $1.04117\pm0.00046$ & $1.04121\pm0.00043$ & $1.04092\pm0.00042$ & $1.04075\pm0.00042$ & $1.04034\pm0.00034$ & $1.04043\pm0.00031$ & $1.04041\pm0.00029$ \\
  $\tau$ & $0.0537\pm0.0079$ & $0.0519\pm0.0073$ & $0.0522\pm0.0076$ & $0.0535\pm0.0080$ & $0.0486\pm0.0077$ & $0.0455^{+0.0083}_{-0.0069}$ & $0.0481\pm0.0075$ \\
  $M$ & $0.914^{+0.009}_{-0.013}$ & $0.919^{+0.017}_{-0.008}$ & $0.8958^{+0.0027}_{-0.0038}$ & $0.8933^{+0.0025}_{-0.0031}$ & $0.9190^{+0.0049}_{-0.0042}$ & $0.9173\pm0.0024$ & $0.9205\pm0.0021$ \\
  ${\rm{ln}}(10^{10}A_s)$ & $3.038\pm0.019$ & $3.032\pm0.017$ & $3.040\pm0.018$ & $3.047\pm0.018$ & $3.047\pm0.017$ & $3.038^{+0.018}_{-0.015}$ & $3.044\pm0.016$ \\
  $n_s$ & $0.9603\pm0.0088$ & $0.9611\pm0.0080$ & $0.9592\pm0.0080$ & $0.9657\pm0.0083$ & $0.9721\pm0.0089$ & $0.9633\pm0.0082$ & $0.9682\pm0.0075$ \\
  $N_{\rm eff}$  & $2.92\pm0.19$ & $2.91\pm0.18$ & $2.97\pm0.18$ & $3.11\pm0.19$ & $3.34\pm0.18$ & $3.18\pm0.14$ & $3.25\pm0.13$ \\
   \hline
  $H_0 {\rm[km/s/Mpc]}$ & $75.9^{+1.4}_{-2.6}$ & $77.0\pm2.6$ & $73.23\pm0.90$ & $73.8^{+0.9}_{-1.0}$ & $75.7\pm1.2$ & $74.80\pm0.86$ & $75.91\pm0.76$ \\
  $\sigma_8$ & $0.890^{+0.015}_{-0.024}$ & $0.895\pm0.021$ & $0.874\pm0.011$ & $0.874\pm0.011$ & $0.9525\pm0.0087$ & $0.9480^{+0.0088}_{-0.0077}$ & $0.9505\pm0.0080$ \\
  $S_8$ & $0.804\pm0.016$ & $0.795\pm0.014$ & $0.824\pm0.014$ & $0.822\pm0.015$ & $0.869\pm0.017$ & $0.875\pm0.011$ & $0.8648\pm0.0096$ \\
  $\Omega_m$ & $0.245^{+0.016}_{-0.009}$ & $0.238^{+0.014}_{-0.019}$ & $0.2667\pm0.0045$ & $0.2657\pm0.0051$ & $0.2500^{+0.0076}_{-0.0085}$ & $0.2556\pm0.0056$ & $0.2484\pm0.0048$ \\
  $r_{\rm drag}$ [Mpc]& $148.4\pm2.0$ & $148.6\pm1.8$ & $147.6\pm1.8$ & $146.3\pm1.8$ & $143.9\pm1.5$ & $145.2\pm1.1$ & $144.7\pm1.0$ \\
 \hline
  $\chi^2_{\rm bf}$ & $2770.330$ & $2778.124$ & $2789.842$ & $3839.774$ & $2770.876$ & $3856.336$ & $2790.254$ \\
 $\Delta \chi^2_{\rm bf}$ & $-2.32$ & $-3.92$ & $+10.13$ & $+32.27$ & $-20.96$ & $+42.16$ & $-6.80$\\    
 \hline
  \hline
\end{tabular}
}
\end{center}
\label{table}
\end{table*}

\subsubsection{VM-VEV+$N_{\rm eff}$}

In this section we discuss another minimal extension of the VM-VEV model where the effective number of neutrinos $N_{\rm eff}$ has been considered along with 7-parameters of the VM-VEV model. The results for different observational datasets are shown in Table~\ref{tab:Mnnu} for some cosmological parameters of interest.
As already noticed in the previous section, in the VM-VEV extension the Hubble constant takes a lower value than the original VM model, more in agreement with R19.

If we look at the constraints for CMB alone, shown in the second column of Table~\ref{tab:Mnnu}, we see that $H_0 = 75.9^{+1.4}_{-2.6}$ km/s/Mpc at 68\% CL, with smaller error bars but the same mean value of the VM-VEV+$\sum m_{\nu}$ model of Table~\ref{tab:Mmnu}. 
The addition of the BAO and Pantheon measurements, instead, contributes to an additional lowering of the Hubble constant, but impressively in agreement with R19 in all the cases.

Moreover, for this VM-VEV extension, the $N_{\rm eff}$ bounds are exactly in agreement with $3.046$ for all the dataset combinations, but when the R19 prior is included. For the latter, because of the correlation between $H_0$ and $N_{\rm eff}$, we find a larger value for $N_{\rm eff}$, showing an indication for a extra radiation at recombination at more than $1\sigma$, for both CMB+R19 and CMB+BAO+R19.


\begin{figure*}
\includegraphics[width=0.72\textwidth]{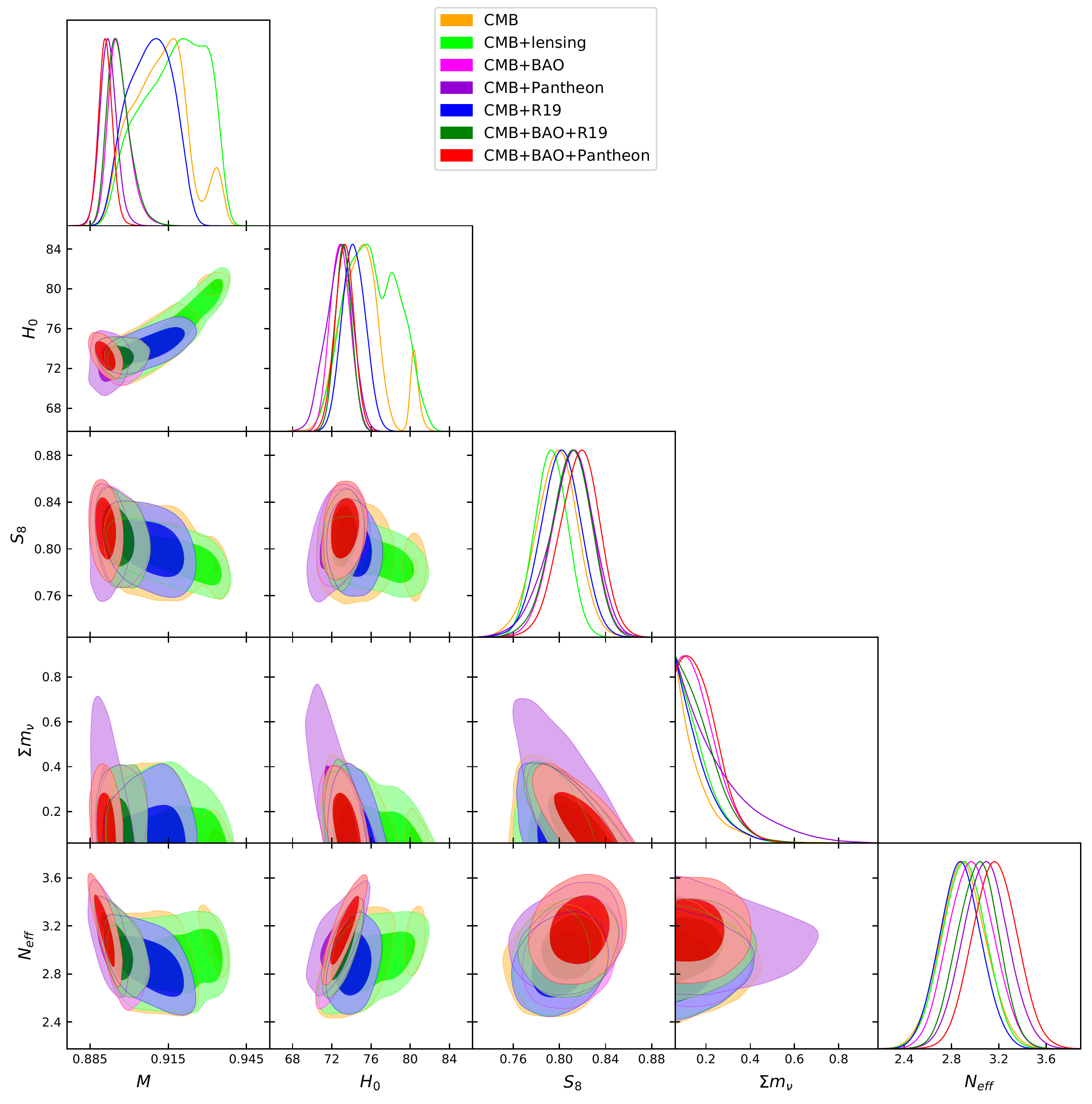}
\caption{68\% and 95\% CL constraints on the VM-VEV+$N_{\rm eff}$+$\sum m_{\nu}$ case. }
\label{fig:Mmnunnu}
\end{figure*}
\begin{table*}[tb]
\caption{ 68\% CL constraints and 95\% CL upper limits on the cosmological parameters of the scenario VM-VEV+$N_{\rm eff}$+$\sum m_{\nu}$ for different dataset combinations explored in this work. The $\Delta \chi^2_{\rm bf}$ (best fit) is relative to the corresponding data best fits within $\Lambda$CDM.
} 
\label{tab:Mmnunnu} 
\begin{center}
\resizebox{\textwidth}{!}{  
\begin{tabular}{ c |c c c c c c c} 
  \hline
 \hline
 Parameters & CMB & CMB+lensing & CMB+BAO & CMB+Pantheon & CMB+R19  & CMB+BAO+Pantheon & CMB+BAO+R19 \\ 

 \hline
  $\Omega_b h^2$ & $0.02222\pm0.00023$ & $0.02225\pm0.00022$ & $0.02222\pm0.00021$ & $0.02230\pm0.00023$ & $0.02219\pm0.00020$ & $0.02237\pm0.00021$ & $0.02227\pm0.00019$ \\
  $\Omega_c h^2$ & $0.1181\pm0.0030$ & $0.1179\pm0.0029$ & $0.1199\pm0.0031$ & $0.1219\pm0.0030$ & $0.1180\pm0.0029$ & $0.1231\pm0.0029$ & $0.1205\pm0.0028$ \\
  $100\theta_{MC}$ & $1.04110\pm0.00045$ & $1.04111\pm0.00044$ & $1.04089\pm0.000344$ & $1.04066\pm0.00043$ & $1.04110\pm0.00044$ & $1.04056\pm0.00041$ & $1.04083\pm0.00041$ \\
  $\tau$ & $0.0534\pm0.0079$ & $0.0531\pm0.0076$ & $0.0532\pm0.0076$ & $0.0542^{+0.0072}_{-0.0081}$ & $0.0535\pm0.0081$ & $0.0541\pm0.0078$ & $0.0537\pm0.0077$ \\
  $M$ & $0.912^{+0.010}_{-0.012}$ & $0.917^{+0.016}_{-0.009}$ & $0.8963^{+0.0030}_{-0.0047}$ & $0.8920^{+0.0029}_{-0.0033}$ & $0.9081^{+0.0093}_{-0.0081}$ & $0.8910\pm0.0027$ & $0.8961^{+0.0031}_{-0.0051}$ \\
  ${\rm{ln}}(10^{10}A_s)$ & $3.037\pm0.019$ & $3.036\pm0.017$ & $3.041\pm0.018$ & $3.048\pm0.018$ & $3.037\pm0.019$ & $3.051\pm0.018$ & $3.044\pm0.018$ \\
  $n_s$ & $0.9588\pm0.0089$ & $0.9595\pm0.0083$ & $0.9594\pm0.0082$ & $0.9632\pm0.0086$ & $0.9575\pm0.0079$ & $0.9661\pm0.0081$ & $0.9617\pm0.0075$ \\
  $\sum m_{\nu} [eV]$  & $<0.365$ & $<0.330$ & $<0.341$ & $<0.565$ & $<0.346$ & $<0.341$ & $<0.352$ \\
  $N_{\rm eff}$  & $2.90\pm0.19$ & $2.91\pm0.18$ & $2.97\pm0.17$ & $3.09\pm0.19$ & $2.88\pm0.18$ & $3.17\pm0.19$ & $3.02\pm0.17$ \\
   \hline
   \hline
  $H_0 {\rm[km/s/Mpc]}$ & $75.2^{+1.6}_{-2.4}$ & $76.2\pm2.6$ & $72.9\pm1.0$ & $72.7^{+1.5}_{-1.2}$ & $74.2\pm1.2$ & $73.28\pm0.95$ & $73.19\pm0.85$ \\
  $\sigma_8$ & $0.871^{+0.029}_{-0.025}$ & $0.878\pm0.025$ & $0.853^{+0.023}_{-0.016}$ & $0.842^{+0.037}_{-0.016}$ & $0.865^{+0.022}_{-0.017}$ & $0.854^{+0.024}_{-0.016}$ & $0.854^{+0.023}_{-0.016}$ \\
  $S_8$ & $0.797\pm0.018$ & $0.793\pm0.014$ & $0.811^{+0.018}_{-0.016}$ & $0.809^{+0.022}_{-0.017}$ & $0.801\pm0.017$ & $0.817^{+0.018}_{-0.016}$ & $0.810^{+0.018}_{-0.016}$ \\
  $\Omega_m$ & $0.252^{+0.016}_{-0.012}$ & $0.246\pm0.016$ & $0.2714^{+0.0054}_{-0.0064}$ & $0.278^{+0.007}_{-0.014}$ & $0.2576\pm0.0095$ & $0.2745^{+0.0050}_{-0.0062}$ & $0.2701^{+0.0053}_{-0.0060}$ \\
  $r_{\rm drag}$ [Mpc] & $148.5\pm1.9$ & $148.5\pm1.9$ & $147.7\pm 1.9$ & $146.5\pm1.8$ & $148.7\pm1.8$ & $145.7\pm1.8$ & $147.2\pm1.7$ \\
   \hline
  $\chi^2_{\rm bf}$ & $2770.556$ & $2778.044$ & $2791.324$ & $3840.180$ & $2769.750$ & $3857.398$ & $2791.562$ \\
 $\Delta \chi^2_{\rm bf}$ & $-2.09$ & $-4.00$ & $+11.61$ & $+32.68$ & $-22.09$ & $+43.22$ & $-5.49$\\   
   
 \hline
  \hline
\end{tabular}
}
\end{center}
\label{table}
\end{table*}
\begin{table*}[tb]
\caption{ 68\% CL constraints and 95\% CL upper limits on the cosmological parameters of the scenario VM-VEV+$N_{\rm eff}$+$\sum m_{\nu}$+$\omk$ for different dataset combinations explored in this work. The $\Delta \chi^2_{\rm bf}$ (best fit) is relative to the corresponding data best fits within $\Lambda$CDM. 
} 
\label{tab:Mmnunnuomk} 
\begin{center}
\resizebox{\textwidth}{!}{  
\begin{tabular}{ c |c c c c c c c} 
  \hline
 \hline
 Parameters & CMB & CMB+lensing & CMB+BAO & CMB+Pantheon & CMB+R19  & CMB+BAO+Pantheon & CMB+BAO+R19 \\ 

 \hline
  $\Omega_b h^2$ & $0.02254\pm0.00026$ & $0.02233\pm0.00023$ & $0.02236\pm0.00023$ & $0.02253\pm0.00026$ & $0.02249\pm0.00024$ & $0.02241\pm0.00025$ & $0.02241\pm0.00022$ \\
  $\Omega_c h^2$ & $0.1185\pm0.0030$ & $0.1176\pm0.0029$ & $0.1193\pm0.0030$ & $0.1213^{+0.0038}_{-0.0033}$ & $0.1193\pm0.0030$ & $0.1229\pm0.0031$ & $0.1201\pm0.0028$ \\
  $100\theta_{MC}$ & $1.04096\pm0.00045$ & $1.04103\pm0.00046$ & $1.04097\pm0.00043$ & $1.04072\pm0.00049$ & $1.04097\pm0.00043$ & $1.04057\pm0.00042$ & $1.04088\pm0.00042$ \\
  $\tau$ & $0.0471\pm0.0080$ & $0.0480\pm0.0086$ & $0.0532\pm0.0078$ & $0.0514\pm0.0081$ & $0.0529\pm0.0077$ & $0.0543^{+0.0073}_{-0.0082}$ & $0.0538\pm0.0079$ \\
  $M$ & $0.75^{+0.12}_{-0.07}$ & $0.869^{+0.064}_{-0.054}$ & $0.9032^{+0.0048}_{-0.0098}$ & $0.893^{+0.024}_{-0.001}$ & $0.9203^{+0.0099}_{-0.0045}$ & $0.8920^{+0.0046}_{-0.0040}$ & $0.9034^{+0.005}_{-0.011}$ \\
  ${\rm{ln}}(10^{10}A_s)$ & $3.024\pm0.019$ & $3.023\pm0.020$ & $3.040\pm0.018$ & $3.041\pm0.020$ & $3.039\pm0.018$ & $3.051\pm0.019$ & $3.044\pm0.018$ \\
  $n_s$ & $0.9695\pm0.0094$ & $0.9639\pm0.0088$ & $0.9638\pm0.0088$ & $0.9705\pm0.0094$ & $0.9681\pm0.0089$ & $0.9676\pm0.0091$ & $0.9662\pm0.0082$ \\
$\sum m_{\nu} [eV]$  & $0.45^{+0.14}_{-0.36}$ & $<0.804$ & $<0.304$ & $<0.559$ & $<0.294$ & $<0.421$ & $<0.280$ \\
  $N_{\rm eff}$  & $3.06\pm0.20$ & $2.96\pm0.19$ & $3.01\pm0.19$ & $3.16\pm0.21$ & $3.06\pm0.19$ & $3.18\pm0.20$ & $3.07\pm0.17$ \\
  $\Omega_k$  & $-0.073^{+0.043}_{-0.019}$ & $-0.021^{+0.020}_{-0.013}$ & $-0.0035^{+0.0029}_{-0.0024}$ & $-0.022^{+0.018}_{-0.005}$ & $-0.0090^{+0.0040}_{-0.0046}$ & $-0.0009^{+0.0022}_{-0.0028}$ & $-0.0034^{+0.0031}_{-0.0024}$ \\
   \hline
   \hline
  $H_0 {\rm[km/s/Mpc]}$ & $50\pm6$ & $65^{+10}_{-9}$ & $72.8^{+1.0}_{-1.1}$ & $66.2^{+5.3}_{-8.5}$ & $73.2\pm1.4$ & $73.14\pm0.98$ & $73.19\pm0.88$ \\
  $\sigma_8$ & $0.715\pm0.069$ & $0.79^{+0.11}_{-0.09}$ & $0.864^{+0.022}_{-0.018}$ & $0.839^{+0.054}_{-0.031}$ & $0.890^{+0.025}_{-0.019}$ & $0.853^{+0.030}_{-0.014}$ & $0.868^{+0.021}_{-0.018}$ \\
  $S_8$ & $1.002\pm0.055$ & $0.846^{+0.048}_{-0.054}$ & $0.820^{+0.020}_{-0.017}$ & $0.890^{+0.041}_{-0.059}$ & $0.840\pm0.026$ & $0.817^{+0.025}_{-0.016}$ & $0.822\pm0.018$ \\
  $\Omega_m$ & $0.61^{+0.11}_{-0.21}$ & $0.366^{+0.10}_{-0.13}$ & $0.2705\pm0.0062$ & $0.342^{+0.008}_{-0.061}$ & $0.267^\pm0.012$ & $0.2755^{+0.0051}_{-0.0064}$ & $0.2690\pm0.0058$ \\
  $r_{\rm drag}$ [Mpc] & $147.1\pm1.9$ & $148.1\pm1.9$ & $147.53\pm 1.9$ & $146.0^{+1.9}_{-2.2}$ & $147.1\pm1.9$ & $145.6\pm1.9$ & $146.9\pm1.7$ \\
   \hline
  $\chi^2_{\rm bf}$ & $2762.914$ & $2776.592$ & $2789.434$ & $3802.992$ & $2765.23$ & $3856.728$ & $2787.794$ \\
 $\Delta \chi^2_{\rm bf}$ & $-9.74$ & $-5.45$ & $+9.72$ & $-4.51$ & $-26.61$ & $+42.55$ & $-9.26$\\  
 
 \hline
  \hline
\end{tabular}
}
\end{center}
\label{table}
\end{table*}
\subsubsection{VM-VEV+$N_{\rm eff}$+$\sum m_{\nu}$}

In this section we discuss the VM-VEV extension for which the total neutrino mass $\sum m_{\nu}$ and the effective number of neutrino species $N_{\rm eff}$ are varying at the same time of the 7-parameters characterizing the VM-VEV model. The results for the different dataset combinations are listed in Table~\ref{tab:Mmnunnu}, and the 1D posterior distributions together with the 2D contour plots are shown in Fig.~\ref{fig:Mmnunnu}.  

The constraints we obtain for the cosmological parameters of interest in this scenario, are very similar to those of the previous two sections. Here, for all the dataset combinations, the Hubble constant is always in agreement with R19 within $1\sigma$, we have only an upper limit for the total neutrino mass, and the effective number of relativistic neutrinos is always in agreement with $3.046$ at 68\% CL. In addition, it is noteworthy that BAO, Pantheon, as well as their combination, are preferring $H_0$ consistent with R19.

In Fig.~\ref{fig:Mmnunnu} we can see that in this VM-VEV extension, all the dataset combinations are in agreement and overlap, solving the disagreement visible in Fig.~\ref{fig:mnunnu} for the corresponding original VM extension with the Pantheon dataset. Moreover, in Fig.~\ref{fig:Mmnunnu} we see that there is a smaller second peak appearing in the CMB only case for $H_0$ and $M$ vacuum, which disappears when more datasets are included in the analysis.

\subsubsection{VM-VEV+$N_{\rm eff}$+$\sum m_{\nu}$+$\omk$}

Finally, we analyse the full extension of the VM-VEV model where the total neutrino mass $\sum m_{\nu}$, the effective number of neutrinos $N_{\rm eff}$, and the curvature energy density $\omk$ are varying freely at the same time of the 7-parameters of the VM-VEV model. All the results of the analysis are shown in Table \ref{tab:Mmnunnuomk}. 
For this model extension, the cosmological parameters have a similar behaviour to that described in Sec.~\ref{omk1}  for the original VM+$N_{\rm eff}$+$\sum m_{\nu}$+$\omk$ case.

In fact, regarding the CMB data only, we have a much lower $H_0 = 50\pm6$ km/s/Mpc (68\% CL), in $3.9\sigma$ tension with R19. Here there is a $1\sigma$ indication for a total neutrino mass different from zero ($\sum m_{\nu}=0.45^{+0.14}_{-0.36}$ eV at 68\% CL), and more than $3\sigma$ evidence for a closed Universe ($\omk = -0.073^{+0.043}_{-0.019}$ at 68\% CL).

We find that, in this scenario, the dataset combinations involving R19 and BAO are not reliable, but shown for completeness. CMB+lensing and CMB+Pantheon can be safely analysed, instead, showing only an upper limit for the total neutrino mass, a neutrino effective number always in agreement with $3.046$ at 68\% CL, and an indication for a closed universe at more than $1\sigma$.

\section{Discussion and Conclusions} 
\label{sec:concl} 

We have studied the impact of extending the neutrino sector beyond the 3 massless SM neutrinos of $\Lambda$CDM in the cosmological evolution of the VM model. This model is known to ameliorate the tension between the observed and $\Lambda$CDM predicted values of the Hubble constant~\cite{DiValentino:2017rcr,DiValentino:2020kha}.

We have shown that for the 8-parameter (VM+$\sum m_\nu$+$N_{\rm eff}$) spatially-flat model and the Planck+BAO+Pantheon dataset combination, there is more than $3\sigma$ evidence for $\sum m_{\nu}=0.80^{+0.18}_{-0.22}~{\rm eV}$ and no indication for extra neutrino species. This combination leads to
$H_0=71.0\pm1.2$~km/s/Mpc, which roughly saturates the $1\sigma$ range lower-boundary of the R19 measurement~\cite{Riess:2019cxk}. The 1D posterior distribution of $H_0$ is compatible with the result of the 6-parameter fit reported in~\cite{DiValentino:2020kha}, but the goodness of the fit is actually slightly improved in the 8-parameter fit when compared to the 6-parameter fit, with $\Delta \chi^2_{\rm bf} = 77.21$ and $\Delta \chi^2_{\rm bf} = 95.83$, respectively. The 8-parameter model when confronted to the Planck+BAO+Pantheon data set combination also yields a best fit-value $S_8=0.755\pm0.032$, addressing the tension with the weak lensing measurements. A statistically significant improvement is observed in the best-fit value of the $S_8$ parameter when compared to the 6-parameter fit, which yields $S_8 = 0.880 \pm 0.010$~\cite{DiValentino:2020kha}. However, when consideration is given to the CMB+Pantheon data sets alone, the best fit $H_0$ value of the 8-parameter model is inconsistent at more than $4\sigma$ with the R19 measurement. This is not the case for the 6-parameter model, which remains consistent with R19 at the $1\sigma$ level when considering the CMB+Pantheon data set. 

We have also shown (see Tables~\ref{tab:Mmnu}, \ref{tab:Mnnu}, and \ref{tab:Mmnunnu}) that the three combinations of spatially-flat VM-VEV models studied herein (i.e., VM-VEV $+ \sum m_\nu$, VM-VEV $+ N_{\rm eff}$, VM-VEV $+ \sum m_\nu + N_{\rm eff}$) can resolve the $H_0$ tension at the $1\sigma$ level, independently of the combination of the selected data samples. Here, the $1\sigma$ significance is defined with respect to R19 observations~\cite{Riess:2019cxk}. However, as shown in Fig.~\ref{fig:Mmnunnu}, the 1D posterior distributions of $H_0$ for CMB and CMB + lensing data sets are multi-peaked. The goodness of the fits for the VM-VEV $+ \sum m_\nu + N_{\rm eff}$ 9-parameter model are comparable to those obtained in~\cite{DiValentino:2020kha} for the 7-parameter VM-VEV model. However, the spatially-flat VM-VEV model endowed with neutrino physics provides a small but statistically significant improvement with respect to the plain VM-VEV model studied in~\cite{DiValentino:2020kha}, for which the fit to CMB + lensing data is only consistent with R19 results at the $2\sigma$ level. All of the 3 VM-VEV-neutrino models provide a solution to the $S_8$ tension with weak lensing measurements.  For $S_8$, all the 1D posterior distributions are single-peaked and for the VM-VEV $+ \sum m_\nu + N_{\rm eff}$ model the 68\% CL regions overlap in a range consistent with local measurements. For the 9-parameter model, there is no indication of extra neutrino species. The 95\% CL limits on the neutrino mass scale $\sum m_\nu$ are slightly less restrictive than those reported by the Planck Collaboration~\cite{Aghanim:2018eyx}.

We have also considered the 9- and 10-parameter model extensions in which the curvature of space $\Omega_k$ is allowed to float as a free parameter in the fit. For the Planck+BAO+Pantheon dataset, both the 9-parameter (VM $+ \sum m_\nu + N_{\rm eff}+ \Omega_k$) and 10-parameter (VM-VEV $+ \sum m_\nu + N_{\rm eff}+ \Omega_k$) models provide a solution to the $H_0$ tension in which the curvature parameter is consistent with that of a spatially-flat universe. Indeed, there is an unnoticeable variation in the rest of the cosmological parameters with respect to the model in which $\Omega_k$ is manually fixed to zero in the fit. In general the datasets favoring a close universe lead to smaller values of $H_0$. The exception is the CMB + BAO data sample which predicts a comparable value of $H_0$, but at the expense of increasing the $S_8$ tension. 

It is important to note that while, in general, the total fit of Planck+BAO+Pantheon is worsening with respect to the $\Lambda$CDM model when the R19 prior is not accounted for, the $\chi^2$ from Planck alone is improved for all of the VM and VM-VEV extensions. This is indicating that the models explored in our paper always provide a better fit of the CMB data. Therefore, even if the joint combination gives a worse $\chi^2$, one should taken into consideration that together with the breaking of the parameters correlation due to the inclusion of additional datasets, each dataset can bring its own systematic errors. Moreover, it should be noticed here that the VM models are not nested with the $\Lambda$CDM one, so there is not a combination of the parameters that can mimic $\Lambda$CDM and its fit. Therefore, in principle, the robustness of the additional datasets from BAO and Pantheon should be tested for these exotic cosmologies, before combining them with Planck data.

In summary, we have demonstrated that the inclusion of beyond SM neutrino physics into the VM+VEV model provides a promising framework to tackle the tensions on both the expansion rate and the clustering of matter. In particular, for all possible combinations of the datasets considered in this paper, the 9-parameter (VM-VEV $+ \sum m_\nu + N_{\rm eff}$) model provides a simultaneous resolution of the $H_0$ and $S_8$ tensions at the $1\sigma$ level. The future CMB-S4 experiment, with a 95\% CL sensitivity to constrain $\Delta N_{\rm eff} \leq 0.06$~\cite{Abazajian:2019eic}, will be able to test the touch of neutrinos on the vacuum metamorphosis.

\bigskip 
\begin{acknowledgments}
The authors thank the referee for some useful comments that helped us to clarify some points. 
EDV acknowledges the support of the Addison-Wheeler Fellowship awarded by the Institute of Advanced Study at Durham University.
SP gratefully acknowledges the Science and Engineering Research Board, Govt. of India, for their Mathematical Research Impact-Centric Support Scheme (File No. MTR/2018/000940).
WY was supported by the National Natural Science Foundation of China under Grants No. 11705079 and No. 11647153. LAA was supported by the U.S. National Science Foundation (NSF
Grant PHY-1620661) and the National Aeronautics and Space
Administration (NASA Grant 80NSSC18K0464).
\end{acknowledgments}

\bibliography{biblio}
\end{document}